\documentclass[final,5p,twocolumn,number]{elsarticle} 

\usepackage{amsmath}
\usepackage{amsfonts}
\usepackage{amssymb}
\usepackage{amsthm}
\usepackage{mathtools}
\usepackage{physics}
\usepackage{siunitx}
\usepackage{optidef}
\usepackage{eurosym}

\usepackage{adjustbox}
\usepackage{epstopdf}
\usepackage{float}
\usepackage{graphicx}
\usepackage{rotating}
\usepackage{subcaption}

\usepackage{tikz}
\usepackage{pgfplots}
\usepackage{pgfplotstable}
\pgfplotsset{compat=1.10}
\usepgfplotslibrary{fillbetween}
\usetikzlibrary{calc}
\usetikzlibrary{positioning}
\usetikzlibrary{decorations.markings}
\usetikzlibrary{trees}
\usetikzlibrary{patterns}
\usetikzlibrary{decorations.pathreplacing}
\usepackage{tikz-qtree}

\usepackage{booktabs}
\usepackage{ctable}
\usepackage{longtable}
\usepackage{threeparttable}
\usepackage{tabularx}
\usepackage{multirow}

\usepackage{xcolor}
\usepackage{color}
\usepackage[ruled]{algorithm2e}
\usepackage[labelfont={bf}]{caption}
\usepackage{enumerate}
\usepackage{lastpage}
\usepackage{listings}
\usepackage[colorlinks=true, allcolors=blue]{hyperref}
\usepackage[title]{appendix}

\usepackage{verbatim}
\usepackage{url}

\newcommand{\E}{\mathbb{E}}

\DeclareMathOperator{\cvar}{CVaR}

\newcommand{\R}{\mathbb{R}}

\date{\today} 
\allowdisplaybreaks[1]

\theoremstyle{definition}

\newtheorem{example}{Example}

\journal{arXiv}

\begin{document}

\begin{frontmatter}

\title{Welfare compensation in international transmission expansion planning under uncertainty}


\author[NTNU]{E. Ruben van Beesten\corref{correspondingauthor}}
\ead{ruben.van.beesten@ntnu.no}
\cortext[correspondingauthor]{Corresponding author}

\author[NTNU]{Ole Kristian Ådnanes}

\author[NTNU]{Håkon Morken Linde}

\author[NTNU]{Paolo Pisciella}

\author[NTNU]{Asgeir Tomasgard}

\address[NTNU]{Norwegian University of Science and Technology (NTNU), Department of Industrial Economics and Technology Management, NO-7491, Trondheim, Norway}

\begin{abstract}
In transmission expansion planning, situations can arise in which an expansion plan that is optimal for the system as a whole is detrimental to a specific country in terms of its expected economic welfare. If this country is one of the countries hosting the planned capacity expansion, it has the power to veto the plan and thus, undermine the system-wide social optimum. To solve this issue, welfare compensation mechanisms may be constructed that compensate suffering countries and make them willing to participate in the expansion plan. In the literature, welfare compensation mechanisms have been developed that work in expectation. However, in a \textit{stochastic} setting, even if the welfare effect after compensation is positive in expectation, countries might still be hesitant to accept the \textit{risk} that the actual, realized welfare effect may be negative in some scenarios.

In this paper we analyze welfare compensation mechanisms in a stochastic setting. We consider two existing mechanisms, lump-sum payments and purchase power agreements, and we develop two novel mechanisms, based on the flow through the new transmission line and its economic value. Using a case study of the Northern European power market, we investigate how well these mechanisms succeed in mitigating risk for the countries involved. Using a theoretically ideal model-based mechanism, we show that there is a significant potential for mitigating risk through welfare compensation mechanisms. Out of the four practical mechanisms we consider, our results indicate that a mechanism based on the economic value of the new transmission line is most promising.
\end{abstract}


\end{frontmatter}

\section{Introduction}
\label{sec:introduction}

The transition from carbon-based electricity production to renewable sources provides pressing arguments for investing in more transmission capacity between European countries \cite{kristiansen2018}. Renewable power sources, such as wind farms and photovoltaic panels are often more efficiently placed at certain geographical locations with, e.g., higher levels of wind or sunshine than others. A well-developed international transmission network can help achieve a system where production can take place where it is opportune, and consumption where it is needed \cite{UN2006multi}. Furthermore, renewable sources typically have higher levels of uncertainty in the associated production levels \cite{konstantelos2014valuation}. International transmission lines can help balance the power production among geographical regions and thus help mitigate production uncertainty through geographical diversification \cite{hasche2010general}.

An important obstacle to achieving the desired interconnected transmission network is the fact that the welfare benefits and costs associated with transmission expansions are often unevenly distributed among countries \cite{mezHosi2016model}. In fact, situations can arise in which an investment in transmission capacity that is beneficial from a Europe-wide system perspective is detrimental to the economic welfare in an individual country. If this negatively affected country is one of the countries hosting the proposed transmission expansion, then it may block the investment and as a consequence, hurt the system as a whole \cite{huppmann2015national}. This issue motivates the need for compensation mechanisms that distribute welfare gains in order to convince all countries to follow through on the transmission expansion plan and thus help achieve the system optimum \cite{olmos2018transmission}.

We consider welfare compensation mechanisms in international transmission expansion planning (TEP) under uncertainty. In the literature, several studies have been performed that determine compensation amounts that should be paid to compensate countries \textit{in expectation} \cite{hogan2018primer,jansen2015alternative,konstantelos2017integrated,kristiansen2018}. However, by restricting attention to expected values, important effects resulting from uncertainty are neglected. The \textit{actual} benefits/costs depend on realization of uncertain elements, such as renewable production levels and electricity prices at different nodes. Hence, if uncertainty is ignored when constructing compensation mechanisms, countries run the risk that in some scenarios, they compensate more than they benefit or are not compensated enough to cover their welfare loss. 

In this paper, we explore the potential of different compensation mechanisms to mitigate the risk associated with investing in a new transmission line. We use a fixed, lump sum payment as a benchmark and investigate whether other compensation mechanisms can perform better in terms of the risk faced by the countries involved as a consequence of the investment in the new transmission line. One alternative mechanism is proposed in the literature \cite{kristiansen2018}, in the form of  power purchase agreements (PPAs). PPAs are contracts that essentially give a certain country a virtual, fixed price at which it can trade in the power market \cite{kristiansen2018}. Deviations of the spot price from this fixed price are then used to determine the compensation amount to be paid or received. We show that in a stochastic setting only one country can receive a PPA. This is in contrast with the deterministic case, in which multiple PPAs can be constructed, one for each country, such that they balance each other exactly.

Besides these two mechanisms from the literature, we also propose two novel compensation mechanisms. We aim to construct compensation mechanisms that achieve risk mitigation by using scenario information to determine the compensation amount. For this purpose, we propose to base the compensations on economic measures related to the economic value of the proposed transmission line investment in the realized scenario. Specifically, we use the amount of flow through the new transmission line and its economic value as measures to base our novel compensation mechanisms on.

We compare the different compensation mechanisms numerically, using a case study of the Northern-European electricity market, focusing on a new transmission line between Norway and Germany. We test the ability of each compensation mechanism to mitigate the risk associated with the transmission expansion investment for each of the affected countries. Here, we consider risk both in terms of the variability of the welfare effect of the investment and in terms of the expected welfare loss as a result of the investment. We run experiments in two settings: a setting with bilateral compensations between Norway and Germany only and a setting with multilateral compensations between all countries that are significantly affected by the proposed investment. 

In both cases, we show that a theoretically ideal mechanism, that equally shares welfare benefits in every scenario, significantly outperforms lump sum payments in terms of its ability to mitigate risk of the countries involved. This demonstrates the potential for risk mitigation through alternative compensation mechanisms. Out of the other compensation mechanisms, our novel value-based mechanism appears to perform best. In particular, it consistently outperforms the lump sum payments. Hence, we show that risk mitigation can actually be achieved by using scenario-dependent compensation mechanisms. Finally, for PPAs, the question of which country receives the PPA turns out to be crucial: a Germany-based PPA performs bad for both Norway and Germany, while a Norway-based PPA shifts risk from Germany to Norway, compared to the lump sum.

The remainder of this paper is structured as follows. In Section~\ref{sec:literature_review} we review the literature on welfare distribution in TEP. Next, in Section~\ref{sec:illustrative_examples} we provide simple, illustrative examples that motivate the need for welfare compensation. In Section~\ref{sec:TEP} we present and solve a transmission expansion model in the Northern-European market. The resulting optimal transmission expansion plan serves as a test environment for our investigation into different welfare mechanisms in Section~\ref{sec:compensation_mechanisms}. In this section, which constitutes the core of the paper, we discuss existing compensation mechanisms, propose several novel mechanisms, and test their performance in the case study from Section~\ref{sec:TEP}. Section~\ref{sec:conclusion} concludes the paper. Finally, Section~\ref{sec:mathematical_model} and Section~\ref{sec:data} in the appendix contain a description of the mathematical model formulation and the data used, respectively, in the TEP model of Section~\ref{sec:TEP}. 

\section{Literature review}
\label{sec:literature_review}

TEP is an active topic of research within the field of operations research \cite{kristiansen2018}. Much of the literature is aimed at developing methods to find good candidates for TEP investments \cite{latorre2003,hemmati2013comprehensive}. Typically, finding an optimal combination of investments is a challenging task, both from a modeling and a computational point of view. One major difficulty is that many relevant parameters, such as demand levels and renewable energy production, may be uncertain, necessitating stochastic models \cite{Zhao2009flexible}. Furthermore, investments are typically of a discrete nature, which introduces the need for integer decision variables \cite{alguacil2003,delatorre2008}. Finally, in order to properly model the effect of the transmission expansion on the power market, explicit modeling of the market participants through equilibrium models may be required \cite{gabriel2012}.  All these factors make TEP a challenging area of research; see \cite{mahdavi2018transmission} for a recent review of the literature on TEP.

In this paper we are not mainly interested in finding the best transmission expansion plan, however, but in making this plan practically \textit{achievable} by constructing welfare-sharing mechanisms that allocate costs and benefits in such a way that all the relevant actors are willing to follow through on the proposed expansion plan. In the literature, most effort in this direction has been spent on creating mechanisms to share the \textit{investment costs} \cite{erli2005transmission,konstantelos2017integrated,nylund2014regional,roustaei2014transmission}. Currently, the most-used cost-sharing strategy is the \textit{equal share principle} \cite{huang2016mind}. According to this principle, the investment costs are split equally between the two countries hosting a new cable. Before the year 2016, all except for two transmission expansion projects in the EU followed this principle \cite{huang2016mind}. Recently, though, there has been a trend towards the so-called \textit{beneficiaries pay principle} \cite{ACER,FERC2012order1000}. According to this principle, each country should pay a share of the investment cost proportional to its benefit from the investment. A specific allocation method that follows this principle is the \textit{net positive benefit differential} \cite{hogan2018primer, konstantelos2017}. The hope is that such a benefit-based allocation method leads to better incentives and ultimately more investments that are beneficial to the system as a whole.

One shortcoming of these cost allocation methods, however, is that they ignore potential \textit{welfare losses} as a result of transmission expansion. As discussed in the introduction, even if the investment costs are zero, some countries might be worse off as a result of transmission expansion, which is especially problematic if one of these countries is hosting the proposed transmission expansion. Recently, a few authors have recognized this problem and proposed methods to share welfare gains and losses \cite{konstantelos2017integrated,kristiansen2018,churkin2019can,churkin2021review}. These authors take a cooperative game theory perspective and propose compensation amounts between countries, based on different conceptions of fairness, e.g., the net postitive benefit differential \cite{konstantelos2017integrated}, the Shapley value \cite{kristiansen2018}, or the nucleolus \cite{churkin2019can}. 

To the best of our knowledge, all papers in this literature consider a single compensation amount based on the \textit{expected} welfare benefits to the various countries. That is, they ignore the possibility of compensations varying per scenario and the resulting potential for risk mitigation, as discussed in the introduction. Rather, most papers seem to implicitly assume a lump sum payment.

One alternative mechanism has been proposed in the literature: power purchase agreements (PPAs) \cite{kristiansen2018}. Such a PPA is a contract based on a fixed price $\pi^{\text{PPA}}$ for purchasing power for each country. The country then pays an amount proportional to the net flow into the country times the difference between the spot price and the predetermined price $\pi^{\text{PPA}}$ to a fund. If the price $\pi^{\text{PPA}}$ would be less beneficial to the country than the actual spot price at which it trades, then the country pays into the fund. In the reverse case it receives money from the fund. The PPA prices $\pi^{\text{PPA}}$ are determined up front such that in expectation, each country receives their ``fair'' share of the total welfare. This PPA-based mechanism does depend on the \textit{actual} behavior of the system (in terms of flows and prices). Hence, this mechanism could potentially be able to mitigate risk resulting from the transmission expansion faced by the relevant countries. Although Kristiansen et al. \cite{kristiansen2018} propose to use PPAs as a welfare compensation mechanism, they don't recognize their potential for risk mitigation. In fact, the authors assume a deterministic setting. In Section~\ref{sec:compensation_mechanisms} we show that the construction of PPAs is fundamentally different in a stochastic setting.

In the remainder of this paper, we contribute to this literature by testing the performance of various compensation mechanisms in terms of their ability to mitigate the risk of investing in a new transmission cable for the countries involved. We consider both mechanisms from the literature (lump sum payments and PPAs) and novel mechanisms (based on the flow through the new transmission line and its economic value).

\section{Illustrative examples} \label{sec:illustrative_examples}

In this section we present two simple, illustrative examples to motivate our study of welfare compensation mechanisms. In Section~\ref{subsec:two-node} we consider a two-node system and show that in the absence of other nodes, there is no need for welfare distribution. In Section~\ref{subsec:three-node} we extend the model to a three-node system and provide an example in which transmission expansion is desirable from a system point of view, but negatively affects one of the hosting countries. This motivates the need for compensation mechanisms studied in this paper.

\subsection{Two-node system: basics} \label{subsec:two-node}

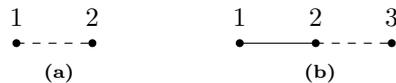
\begin{figure}[h]
    \centering
    \begin{subfigure}[b]{0.25\columnwidth}
        \centering
        \begin{tikzpicture}[scale=0.5]
    \coordinate (n1) at (0,0);
    \coordinate (n2) at (2,0);
    \draw[dashed] (n1) -- (n2);
    \fill[black] (n1) circle (3pt);
    \fill[black] (n2) circle (3pt);
    \node[above=1mm of n1] {1};
    \node[above=1mm of n2] {2};
\end{tikzpicture}
        \caption{}
        \label{subfig:network_configurations_2_nodes}
    \end{subfigure}
    \begin{subfigure}[b]{0.5\columnwidth}
        \centering
        \begin{tikzpicture}[scale=0.5]
    \coordinate (n1) at (0,0);
    \coordinate (n2) at (2,0);
    \coordinate (n3) at (4,0);
    \draw (n1) -- (n2);
    \draw[dashed] (n2) -- (n3);
    \fill[black] (n1) circle (3pt);
    \fill[black] (n2) circle (3pt);
    \fill[black] (n3) circle (3pt);
    \node[above=1mm of n1] {1};
    \node[above=1mm of n2] {2};
    \node[above=1mm of n3] {3};
\end{tikzpicture}
        \caption{}
        \label{subfig:network_configurations_3_nodes}
    \end{subfigure}
    \caption{Network configurations used in the examples in Section~\ref{sec:illustrative_examples}.}
    \label{fig:network_configurations}
\end{figure}

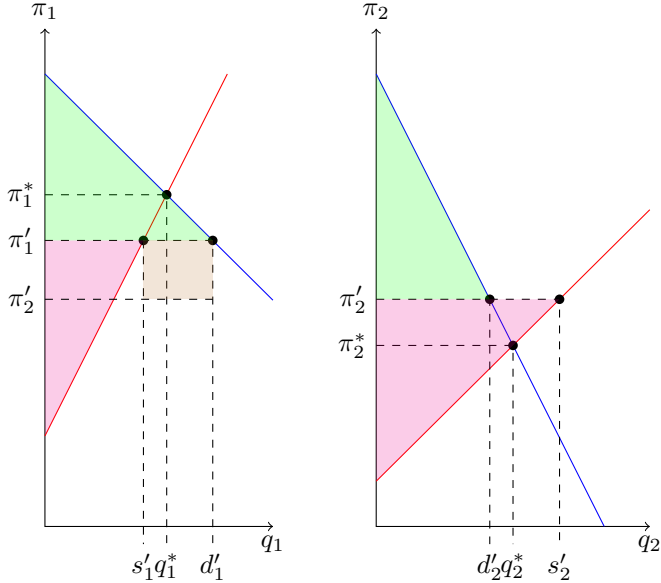
\begin{figure}[t]
    \centering
    \begin{tikzpicture}[scale=0.6]
    \draw [<->] (0,11) node (yaxis) [above] {$\pi_1$} |- (5,0) node (xaxis) [below] {$q_1$};
    \draw[domain=0:5,smooth,variable=\x,blue] plot ({\x},{10-\x});
    \draw[domain=0:4,smooth,variable=\x,red] plot ({\x},{2+2*\x});
    \coordinate (uncap) at (2.67,7.33);
    \fill[black] (uncap) circle (3pt);
    \draw[dashed] (yaxis |- uncap) node[left] {$\pi_1^*$} -| (xaxis -| uncap) node[below] {$q_1^*$};
    \coordinate (capdem) at (3.68,6.32);
    \coordinate (capsup) at (2.16,6.32);
    \coordinate (p2) at (3.68,5.02);
    \fill[black] (capdem) circle (3pt);
    \draw[dashed] (yaxis |- capsup) node[left] {$\pi_1^\prime$} -| (xaxis -| capdem) node[below] {$d_1^\prime$};
    \fill[black] (capsup) circle (3pt);
    \draw[dashed] (capsup |- capdem) node[left] {} -| (xaxis -| capsup) node[below] {$s_1^\prime$};
    \fill[green,opacity=0.2] plot coordinates {(0,10)(0,6.32)(3.68,6.32)};
    \fill[magenta,opacity=0.2] plot coordinates {(0,6.32)(2.16,6.32)(0,2)};
    \fill[brown,opacity=0.2] plot coordinates {(2.16,6.32)(3.68,6.32)(3.68,5.02)(2.16,5.02)};
    \draw[dashed] (0,5.02) node [left] {$\pi_2^\prime$} -- (3.68,5.02);
\end{tikzpicture}
\quad
\begin{tikzpicture}[scale=0.6]
    \draw [<->] (0,11) node (yaxis) [above] {$\pi_2$} |- (6,0) node (xaxis) [below] {$q_2$};
    \draw[domain=0:5,smooth,variable=\x,blue] plot ({\x},{10-2*\x});
    \draw[domain=0:6,smooth,variable=\x,red] plot({\x},{1+\x});
    \coordinate (uncap) at (3,4);
    \fill[black] (uncap) circle (3pt);
    \draw[dashed] (yaxis |- uncap) node[left] {$\pi_2^*$} -| (xaxis -| uncap) node[below] {$q_2^*$};
    \coordinate (capdem) at (2.49,5.02);
    \coordinate (capsup) at (4.02,5.02);
    \fill[black] (capdem) circle (3pt);
    \draw[dashed] (yaxis |- capdem) node[left] {$\pi_2^\prime$} -| (xaxis -| capdem) node[below] {$d_2^\prime$};
    \fill[black] (capsup) circle (3pt);
    \draw[dashed] (capdem |- capsup) node[left] {} -| (xaxis -| capsup) node[below] {$s_2^\prime$};
    \fill[green,opacity=0.2] plot coordinates {(0,10)(0,5.02)(2.49,5.02)};
    \fill[magenta,opacity=0.2] plot coordinates {(0,5.02)(4.02,5.02)(0,1)};
\end{tikzpicture}
    \caption{Surpluses in the two-node example in Section~\ref{subsec:two-node}.}
    \label{fig:surplus_two_nodes}
\end{figure}

First we consider a network consisting of two country nodes, as illustrated in Figure~\ref{subfig:network_configurations_2_nodes}. Each node $i=1,2$ contains power suppliers and consumers, represented by a linear supply and demand curve $S_i(\pi_i)$ and $D_i(\pi_i)$, respectively. We assume perfect competition with all market participants being price takers. Moreover, we assume that that initially, there is no transmission capacity between the two nodes. Hence, initially every node is an independent market and initial equilibrium prices $\pi^*_1$ and $\pi^*_2$ occur where the supply and demand curves meet. See Figure~\ref{fig:surplus_two_nodes} for an illustration. The green and red areas represent the consumer surplus (CS) and producer surplus (PS), respectively, in each node. Note that $\pi^*_1 > \pi^*_2$, indicating that power is more scarce in node 1 as compared to node 2.

\begin{figure}[h]
    \centering
    \begin{tikzpicture}[scale=0.6]
    \coordinate (o) at (0,0);
    \coordinate (xmax) at (4,0);
    \coordinate (ymax) at (0,8);
    \draw[->] (o) -- (xmax);
    \node[right=0mm of xmax]{$f$};
    \draw[->] (o) -- (ymax);
    \node[above=0mm of ymax]{$\pi$};
    
    \draw[domain=0:4,smooth,variable=\x,blue] plot ({\x},{7.33-0.67*\x});
    \coordinate (p1star) at (0,7.33);
    \node[left=0mm of p1star]{$\pi_1^*$};
    
    \draw[domain=0:4,smooth,variable=\x,red] plot ({\x},{0.67*\x+4});
    \coordinate (p2star) at (0,4);
    \node[left=0mm of p2star]{$\pi_2^*$};
    
    \coordinate (kopt) at (1.53,0);
    \draw[dashed] (kopt) -- (1.53,6.32);
    \node[below=0mm of kopt]{$x^\prime$};
    
    \coordinate (p1) at (0,6.32);
    \coordinate (p2) at (0,5.02);
    \coordinate (opt1) at (1.53,6.32);
    \coordinate (opt2) at (1.53,5.02);
    \coordinate (pnew) at (0,5.67);
    \coordinate (optnew) at (2.5, 5.67);
    \draw[dashed] (p1) -- (opt1);
    \node[left=0mm of p1]{$\pi_1'$};
    \fill[black] (opt1) circle (3pt);
    \draw[dashed] (p2) -- (opt2);
    \node[left=0mm of p2]{$\pi_2'$};
    \fill[black] (opt2) circle (3pt);
    \draw[dashed] (pnew) -- (optnew);
    \node[left=1mm of pnew]{$\bar{\pi}$};
    
    \fill[green,opacity=0.2] plot coordinates {(0,7.33)(opt1)(p1)};
    
    \fill[magenta,opacity=0.2] plot coordinates {(0,4)(opt2)(p2)};
    
    \fill[brown,opacity=0.2] plot coordinates {(p1)(opt1)(opt2)(p2)};
    
    \coordinate (opt) at (2.5,8);
    \coordinate (kmax) at (2.5,0);
    \draw[dashed] (kmax) -- (opt);
    \node[below=0mm of kmax]{$\bar{f}$};
\end{tikzpicture}
    \caption{Import/export graph for the two-node example in Section~\ref{subsec:two-node}.}
    \label{fig:intput_output}
\end{figure}
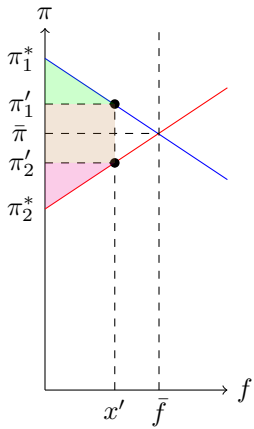

Before introducing the possibility of transmission expansion, it is useful to define the import/export curve corresponding to node 1 and 2. We define the import curve for node 1 as $I_1(\pi_1) := D_1(\pi_1) - S_1(\pi_1)$ and the export curve for node 2 as $E_2(\pi_2) := S_2(\pi_1) - D_2(\pi_2)$. Here, $I_1(\pi_1)$ represents the amount that node 1 is willing to import at price $\pi_1$, while $E_2(\pi_2)$ is the amount that node 2 is willing to export at price $\pi_2$. See Figure~\ref{fig:intput_output} for an illustration of the import/export graph corresponding to the example in Figure~\ref{fig:surplus_two_nodes}. On the horizontal axis we have the variable $f$, representing flow from node 2 to 1 (i.e., import to 1/export from 2). Observe that, by definition, $I_1(\pi^*_1) = 0$ and $E_2(\pi^*_2) = 0$. Moreover, the figure shows that if transmission capacity between node 1 and 2 were unlimited, the combined market would clear at a common price $\pi_1 = \pi_2 = \bar{\pi}$ with an associated flow of $\bar{f}$. 

Now consider a social planner that can invest in transmission capacity $x$ between node 1 and 2. We assume that transmission capacity $x$ can be any non-negative real number and that the marginal cost of investment $C$ is constant. For a given investment $x$, the transmission system operator (TSO) will earn congestion rent: the TSO buys power at the low-price node and sells at the high-price node. It earns an amount $CR = (\pi_1 - \pi_2) f$, where $f$ again represents the flow from node 2 to 1. If $x \geq \bar{f}$, then the transmission capacity constraint $f \leq x$ is be non-binding and we obtain the common price $\bar{\pi}$ and flow $\bar{f}$ described above. Now suppose that $x < \bar{f}$. Then, the transmission capacity constraint is binding and a price difference will arise between the two nodes. An illustration is given in Figures~\ref{fig:surplus_two_nodes} and \ref{fig:intput_output} for a capacity of $x^\prime$. The associated prices, demand and supply levels, and flow are denoted by a prime. 

The welfare gains associated with this capacity are visualized in Figure~\ref{fig:surplus_two_nodes}. The welfare gain to node 1 is represented by the triangle defined by the three black dots in the left graph, and similarly for node 2. The congestion rent is also part of the welfare gain, and is represented by the rectangle in the left graph. Note that it is not necessarily allocated to a specific region, although it is shown in the diagram representing node 1 in the picture. Equivalently, the welfare effects are visualized in Figure~\ref{fig:intput_output}. Here, the upper triangle represents the welfare gain to node 1 and has the same area as the left triangle in Figure~\ref{fig:surplus_two_nodes}. The lower triangle represents the welfare gain to node 2 and has the same area as the right triangle in Figure~\ref{fig:surplus_two_nodes}. Finally, the rectangle represents the congestion rent and has the same size as the rectangle in Figure~\ref{fig:surplus_two_nodes}. 

Figure~\ref{fig:intput_output} reveals how much a welfare-maximizing social planner should invest. The welfare gains are represented by the colored trapezoid. Clearly, the marginal welfare gains are decreasing in $x$. The marginal cost is constant at $C$. If $\pi_1^* - \pi_2^* \leq C$, then an investment of $x=0$ is optimal. In the more interesting case where $\pi_1^* - \pi_2^* > C$, it is optimal to invest an amount $x$ such that $\pi_1 - \pi_2 = C$, and congestion rent equals investment cost. Hence, as expected from economic theory, in the social optimum marginal revenue equals marginal cost. 

Looking at the welfare effects of transmission expansion, we see that in each node there are winners and losers. In node 1 consumers benefit from the lower price, while producers are hurt. In node 2 the effects are reversed. However, it is clear from Figure~\ref{fig:intput_output} that the total welfare effects of transmission expansion, represented by the areas of the two triangles, are non-negative for both nodes. In the next subsection we will see that the latter result need not hold for systems with more than two nodes.

\subsection{Three-node system: welfare issues} \label{subsec:three-node}

Next, we consider a network consisting of three nodes, illustrated in Figure~\ref{subfig:network_configurations_3_nodes}. The purpose is to provide an example in which transmission expansion would be beneficial for the three-node system as a whole, but one of the nodes would be worse off than without the added transmission capacity. In particular, we show that this can be the case for one of the nodes that is hosting the new transmission capacity, which means that they will be able to block the transmission expansion and thus, hurt the system as a whole.

The three-node example we consider is designed to be the simplest possible example that manifests the welfare loss problem outlined in the previous paragraph. Let node 1 be a supply node, meaning that it has only supply, represented by a linear supply curve, and no demand. In contrast, let node 2 and 3 be demand nodes, meaning that they have only demand, represented by a linear demand curve, and no supply. The supply and demand curves are given by $S_1(\pi_1) = \pi_1$, $D_2(\pi_2) = 6 - \pi_2$, $D_3(\pi_3) = 6 - \pi_3$. As in the two-node setting, we assume perfect competition and price-taking behavior for all market participants.

\begin{figure}[t]
    \centering
    \begin{tikzpicture}[scale=0.6]
    \draw [<->] (0,8) node (yaxis) [above] {$\pi$} |- (8,0) node (xaxis) [below] {$f$};
    \draw[domain=0:6,smooth,variable=\x,blue] plot ({\x},{6 - \x});
    \node at (6,1.5) {\textcolor{blue}{$I_2(\pi)$}};
    \draw[domain=0:6,smooth,variable=\x,red] plot({\x},{\x});
    \node at (6,4.5) {\textcolor{red}{$E_1(\pi)$}};
    \coordinate (opt) at (3,3);
    \fill[black] (opt) circle (3pt);
    \draw[dashed] (yaxis |- opt) node[left] {$\pi^{\text{old}}$} -| (xaxis -| opt) node[below] {$f^{\text{old}}$};
    \fill[green,opacity=0.2] plot coordinates {(0,6)(0,3)(3,3)};
    \fill[magenta,opacity=0.2] plot coordinates {(0,0)(0,3)(3,3)};
\end{tikzpicture}
\quad
\begin{tikzpicture}[scale=0.6]
    \draw [<->] (0,8) node (yaxis) [above] {$\pi$} |- (8,0) node (xaxis) [below] {$f$};
    \draw[domain=0:6,smooth,variable=\x,blue] plot ({\x},{6 - 0.5*\x});
    \node at (6.5,2.5) {\textcolor{blue}{$I_{2,3}(\pi)$}};
    \draw[domain=0:6,smooth,variable=\x,red] plot({\x},{\x});
    \node at (6.5,5) {\textcolor{red}{$E_1(\pi)$}};
    \coordinate (optnew) at (4,4);
    \fill[black] (optnew) circle (3pt);
    \draw[dashed] (yaxis |- optnew) node[left] {$\pi^{\text{new}}$} -| (xaxis -| optnew) node[below] {$f^{\text{new}}$};
    \fill[green,opacity=0.2] plot coordinates {(0,6)(0,4)(4,4)};
    \fill[magenta,opacity=0.2] plot coordinates {(0,0)(0,4)(4,4)};
\end{tikzpicture}
    \caption{Import/output graph for the three-node example in Section~\ref{subsec:three-node}}
    \label{fig:surplus_three_nodes}
\end{figure}
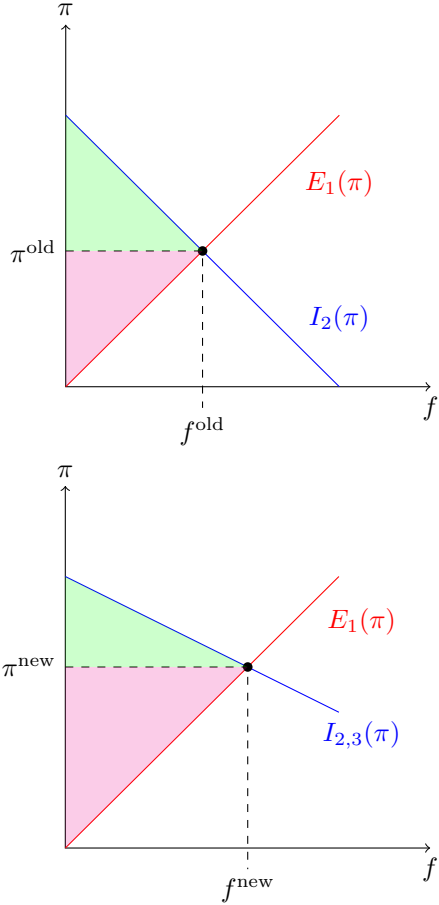

We consider an initial situation in which there is a transmission line between node 1 and 2 of unlimited capacity and no transmission lines between other pairs of nodes. In this situation, node 1 and 2 form a single market with a common price $\pi$ and an equilibrium is found where the supply curve of node 1 meets the demand curve of node 2: $\pi_1 = \pi_2 = \pi^* = 3$, $s_1^* = d_2^* = q^* = 3$. This situation is illustrated in the top graph in Figure~\ref{fig:surplus_three_nodes}. Note that the import/export graphs are given by $I_2 = D_2$ and $E_1 = S_1$. The green area represents the consumer surplus in node 2, while the red are represents the producer surplus in node 1. The welfare distribution in this old situation is described in the top half of Table~\ref{tab:welfare_distribution}.

Next, suppose there is the possibility of opening a transmission line between node 2 and 3. Suppose the cost of building this line is zero and it has unbounded capacity. If the line is built, then node 2 and 3 can be seen as a single demand node and together, node 1, 2 and 3 form a single, joint market, without any constraints on transmission between nodes. Hence, we can find the market equilibrium by aggregating the two demand curves to obtain the joint demand curve $D(\pi) = D_2(\pi) + D_3(\pi) = 12 - 2\pi$. The new equilibrium is found at $\pi_1 = \pi_2 = \pi_3 = \pi^* = 4$, and $s_1 = 2 d_2 = 2 d_3 = q^* = 4$. The new situation is illustrated in the bottom graph in Figure~\ref{fig:surplus_three_nodes}. Note that the import/export graphs are given by $I_{2,3} = D$ and $E_1 = S_1$. The joint consumer surplus in node 2 and 3 is represented by the green area, while the producer surplus in node 1 is given by the red area. 

The welfare distribution in the new situation is described in Table~\ref{tab:welfare_distribution}. Note that the total welfare of the entire system has increased. This is as expected, since the new situation has fewer (transmission) constraints than the old situation. However, observe that the total welfare of node 2 has decreased. The new connection with the demand node 3 has made power scarcer and thus, increased the price. This higher price hurts the consumers in node 2, while there are no producers in node 2 to benefit from the higher price. Hence, node 2 is worse off with the new connection than without. Importantly, node 2 is one of the two end points of the cable and hence, it can be expected to be able to block the connection. Moreover, in practice, a direct connection between node 1 and 3 might be infeasible (or very expensive) for geographical reasons.

In this situation, node 1 and 3 may decide to spend part of their welfare gains to compensate node 2 for its welfare loss. Indeed, they have sufficient welfare gains (3.5 for node 1 and 2 for node 3) to compensate the welfare loss of 2.5 in node 2. Importantly, such a compensation may help achieve the social optimum, which is desirable from a system point of view. This example motivates our investigation into welfare compensation schemes in the remainder of this paper.

\begin{table}[t]
    \centering
    \begin{tabular}{@{}llrrrr@{}}
    \toprule
    \textbf{}  & \textbf{} & \textbf{1} & \textbf{2} & \textbf{3} & \textbf{System} \\ \midrule
    \multirow{3}{*}{Old situation} & CS & 0 & 4.5 & 0 & 4.5 \\
    & PS & 4.5 & 0 & 0 & 4.5 \\ \cmidrule(l){2-6} 
    & TW & 4.5 & 4.5 & 0 & 9 \\ \midrule
    \multirow{3}{*}{New situation} & CS & 8 & 0 & 0 & 8  \\
    & PS & 0 & 2 & 2 & 4 \\ \cmidrule(l){2-6} 
    & TW & 8 & 2 & 2 & 12 \\ \bottomrule
    \end{tabular}%
    \caption{Welfare distribution -- in terms of consumer surplus (CS), producer surplus (PS), and total surplus (PS) -- in the three-node example of Section~\ref{subsec:three-node}.}
    \label{tab:welfare_distribution}
\end{table}

\section{Transmission expansion planning}
\label{sec:TEP}

In this section we use a TEP model to find a system-optimal transmission expansion plan in a case study of the Northern European power market. The results from this model are used in Section~\ref{sec:compensation_mechanisms} to investigate the performance of various welfare compensation mechanisms. The goal here is not to find the best possible transmission expansion plan in real life; more sophisticated models exist in the literature that are likely more capable for that purpose (see, e.g., \cite{mahdavi2018transmission}). Instead, the goal is to find a reasonable candidate transmission expansion plan that can serve as a basis for an analysis of different welfare compensation mechanisms.

The mathematical model can be described as a mathematical program with equilibrium constraints (MPEC), in which a social planner determines the optimal transmission expansion plan from a system welfare point of view, while taking the optimal subsequent behavior of all market participants into account through optimality conditions in the form of equilibrium constraints. A full description of the mathematical model is given in Section~\ref{sec:mathematical_model} in the appendix.

\begin{figure}[t]
\centering
\input{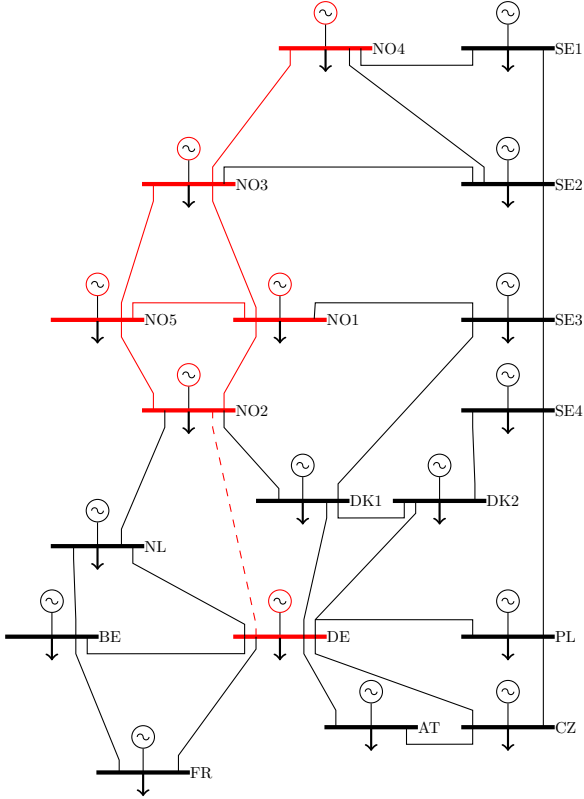}
\caption{Line diagram of the power system in the case study.}
\label{fig:LineDiagram}
\end{figure}

Our mathematical model is used to analyze a case study of the Northern European power market, focusing on possible investment in a new transmission line between the price zones NO2 and DE, representing part of Norway and the whole of Germany, respectively. Figure~\ref{fig:LineDiagram} provides a schematic overview of the power system used in the case study. We allow for investments in transmission cables within Norway, too (i.e., between all zones colored red). However, we disallow investments in any other cables, in order to be able to isolate the effects of investments in the new NO2-DE line.

The data used to parametrize the model is based on historical data on investments in and operation of the Northern European power market. A detailed description of the data used in the case study can be found in Section~\ref{sec:data} in the appendix.

\subsection{Results}
\label{subsec:results}
We first consider a benchmark setting in which we disallow investment in the NO2-DE line, but allow for investments within Norway. In this benchmark a social planner invests in 221 MW of extra capacity in the NO1-NO2 line. This comes at an annualized investment cost of 8.0 million euros.

Next, we allow investment in the NO2-DE line, as well as the other lines within Norway. We find that a social planner would invest in a capacity expansion of 4147 MW in the NO2-DE cable. This comes at an annualized investment cost of 294 million euros per year. As expected, this is equal to the expected yearly congestion rent earned on the new line. Besides investment in the NO2-DE line, a social planner would also invest in 464 MW of extra capacity the NO1-NO2 cable, at an annual investment cost of 16.9 million euros. Note that this investment is larger than in the benchmark setting. The rationale behind this additional investment is to export the electricity produced in NO1 (or imported into NO1) through NO2 to DE and the rest of Europe.

Compared to the benchmark, the social planner solution leads to a system-wide social welfare increase of 84.6 million euros per year. This constitutes a return on investment of 27.9\%. Hence, from a system point of view, it is highly desirable to invest in a new transmission cable. However, some countries -- Germany in particular -- do not profit from the extra capacity and might block the investment plan.

\begin{figure}[h]
    \centering
    \includegraphics[scale=0.75]{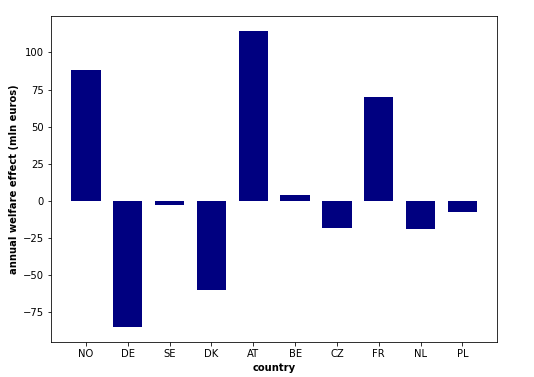}
    \caption{Total welfare effects of the NO2-DE cable per country}
    \label{fig:welfare_country}
\end{figure}

The welfare effects of the transmission investment are illustrated in Figure~\ref{fig:welfare_country}. Three countries benefit significantly from the investment: Norway, Austria, and France. In Norway, producers benefit from their additional capacity to export electricity to mainland Europe, while in Austria and France, consumers benefit from the resulting lower prices. On the other hand, two countries are significantly negatively affected: Germany and Denmark. In both countries, producers suffer from the lower prices as a consequence of cheap Norwegian electricity entering the market. This price drop results in both lower profit margins and loss of sales: some of the demand in Germany will be satisfied through Norwegian power entering the country through the new NO2-DE line. One benefit of this, however, is that the power produced in Norway is more than 95\% renewable \cite{norwayministryofpetroliumenergy2016}, while in Germany it is only about 45\% \cite{germanumweltsbundesamt2021}. Hence, the new line can indirectly contribute to German sustainability goals. 

Nevertheless, the negative welfare impact of the capacity expansion in Germany may well be an obstacle for realization of the transmission expansion. Since Germany is one of the hosting countries, it can block the investment if it deems the new line to be detrimental to its welfare. However, since the system welfare effects are positive, it is in principle  possible to construct compensation mechanisms that result in a net welfare gain for Germany. Notably, since the net welfare gain in Norway (88.2 million annually)  exceeds the net welfare loss in Germany (85.1 million annually), a bilateral compensation scheme between these two countries is a viable option.



\section{Compensation mechanisms}
\label{sec:compensation_mechanisms}

In this section we investigate different compensation mechanisms that may be used to compensate countries for welfare losses resulting from transmission expansion investments. First, in Section~\ref{subsec:incentive_effects} we shortly discuss potential incentive effects of compensation mechanisms. Next, in Section~\ref{subsec:mechanisms} we list and discuss a number of existing compensation mechanisms and we propose two novel mechanisms. We discuss the rationale behind them and their expected performance in a stochastic setting. Finally, in Section~\ref{subsec:performance} we numerically test the performance of the different mechanisms using the case study from Section~\ref{sec:TEP}.

\subsection{Incentive effects}
\label{subsec:incentive_effects}

One important potential effect of compensation mechanisms is that they might skew incentives of actors in the power market. That is, compensation mechanisms may create incentives for certain actors in the power market to change their behavior in order to receive a larger compensation amount. Importantly, such incentive distortions deviate from ``pure'', market-based incentives and hence, may steer the market equilibrium away from a perfect market equilibrium. This may well have negative effects on the total welfare in the system as a whole. In this paper, we steer away from this issue as much as possible, but a short discussion is in place.

In practice, the question whether compensation mechanisms affect incentives depends on \textit{who receives the compensation}. However, in the literature this issue seems to have been ignored. Welfare compensations are modeled as transfers of money between \textit{countries}, but what agent within a country should receive the money is typically not specified. Given the fact that the proposed compensations are meant for compensating the \textit{total welfare} in a country, it seems most reasonable to assume that they are paid between \textit{governments}, which represent the countries as a whole. Since governments do not trade in power markets directly, it seems to be safe to assume that compensations between governments do not skew incentives of any market players. 

However, in practice, governments may decide to use the compensations to, e.g., change taxes/subsidies on power, in order to compensate the groups within the country that are affected by the transmission line investment (e.g., consumers or producers). In this case, the compensation may skew incentives of market players. Moreover, if the tax/subsidy change \textit{depends on the actual amount of compensation received} by the government, then not only the expected amount of compensation, but also the particular compensation \textit{mechanism} may skew incentives of market players. 

In this paper, we assume that governments do not use compensations in a way that affects the incentives of players in the energy market. This assumption is in line with the paradigm of avoiding protection and aiming for perfect competition, on which the European power market is founded \cite{olmos2018transmission}. It allows us to steer away from the issue of incentive distortions as much as possible and investigate the \textit{risk} effects of compensation mechanisms in isolation. We believe that the topic of incentive distortions resulting from compensation mechanisms is a complicated and interesting topic in its own right and deserves attention in future research.

\subsection{Mechanisms}
\label{subsec:mechanisms}

\subsubsection{Lump sum payment: issues}

The most straightforward compensation mechanism is a lump sum payment. This consists of a fixed payment from one country to another. It has the benefits of being very simple and completely predictable. However, in the presence of uncertainty, a lump sum has some drawbacks. In a stochastic setting, the \textit{actual} welfare effect resulting from investment in the new line is uncertain. Hence, the lump sum payment should be based on the \textit{expected} welfare effect. However, there might well be a discrepancy between this expected welfare effect and its actual, realized value. As a result, there may be scenarios in which the lump sum compensation to Germany is not enough or in which Norway must compensate Germany, even though in reality it does not profit from the new line. The potential of such scenarios might make countries hesitant to accept the lump sum mechanism.

More generally speaking, uncertainty about the actual welfare effects introduces \textit{risk} for the countries involved. A lump sum mechanism ignores this risk completely by focusing on only the expected value. Other mechanisms might be able to deal with risk in a smarter way. Ideally, a mechanism compensates countries more in scenarios in which they are hurt more by the new line and vice versa. This would reduce the risk of the countries involved, potentially making them more willing to accept the compensation mechanism.

\begin{figure}[t]
    \centering
    \includegraphics[scale=0.4]{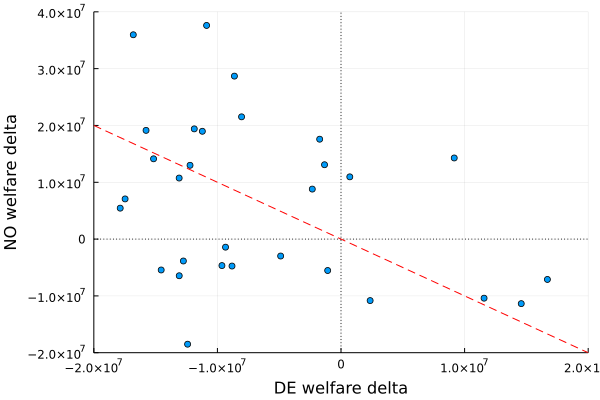}
    \caption{Scatter plot of the effect of the new line on the welfare in Norway and Germany. The dashed red line represents all points for which the aggregated welfare effect for Norway and Germany is zero.}
    \label{fig:scatter_NO_DE_welfare_delta}
\end{figure}

In order to be able to construct a risk-mitigating mechanism, a few conditions need to hold. Firstly, the new line's welfare effects in the compensating countries and compensated countries should be negatively correlated, such that they can share their risk between scenarios. For the relevant countries in our case study, Norway and Germany, this correlation turns out to be $-0.34$; see Table~\ref{tab:corr_coalition}. This moderately negative correlation suggests that there is indeed a potential for risk sharing, although probably only to a modest degree. The relationship between the welfare effects in Norway and Germany is presented in more detail in the scatter plot in Figure~\ref{fig:scatter_NO_DE_welfare_delta}. Note that just over half of the dots, representing scenarios, are to the top-right of the red 45 degree line. In those cases, there is enough total benefit in both countries combined that could theoretically be shared such that neither country suffers from the new line. For the other half of the scenarios this is not possible, however. This again suggests that there is indeed room for risk sharing, although to a limited degree.

Secondly, for a compensation mechanism to be able to exploit such a negative correlation, it should \textit{depend on the scenario}. That is, the compensation amount should be different under different circumstances. Evidently, a lump sum payment lacks this property, so alternatives are warranted.

\begin{table}[]
\centering
\caption{Correlations between the welfare effects of the new transmission line in different countries.}
\label{tab:corr_coalition}
\begin{tabular}{llllll}
\toprule
            & \textbf{NO} & \textbf{AT} & \textbf{FR} & \textbf{DE} & \textbf{DK} \\ \midrule
\textbf{NO} & 1.00        & 0.29        & 0.23        & -0.34       & -0.63       \\
\textbf{AT} & 0.29        & 1.00        & 0.98        & -0.27       & -0.63       \\
\textbf{FR} & 0.23        & 0.98        & 1.00        & -0.31       & -0.57       \\
\textbf{DE} & -0.34       & -0.27       & -0.31       & 1.00        & 0.36        \\
\textbf{DK} & -0.63       & -0.63       & -0.57       & 0.36        & 1.00        \\ \bottomrule
\end{tabular}%
\end{table}

\subsubsection{Power purchase agreement}
One scenario-dependent compensation mechanism proposed in the literature \cite{kristiansen2018} is a so-called power purchase agreement (PPA). This entails giving a certain country, say country A, a virtual, fixed price $\pi^{\text{PPA}_A}$ for importing/exporting power through the new transmission line. Then, after trading at the spot price, the country is compensated such that it is as though the country had traded at the PPA price $\pi^{\text{PPA}_A}$. Mathematiaclly, we define the compensation to country $A$ in scenario $\omega$ by 
\begin{align*}
    C_{A}^{\text{PPA}_A,\omega} = \sum_{t \in \mathcal{T}} f_{AB}^{\omega,t} (\pi^{\text{PPA}_A} - \pi_A^{\omega,t}).
\end{align*}
Assuming country $A$ tends to export (i.e., typically $f_{AB}^{\omega,t}$ is positive), a low PPA price yields a negative compensation to country A, while a high PPA price yields a positive compensation. The reverse holds if country $A$ tends to import. More general definitions of PPAs exist, which give a country a virtual, fixed price for its trade through \textit{all} transmission cables it is connected to \cite{kristiansen2018}. However, we find it unreasonable to expect that a country would be willing to essentially take over all of another country's price risk in a compensation scheme for a single transmission line investment. Hence, we choose to restrict the definition to focus on trade through the new transmission line only.

One important property of a PPA is that it is connected to a specific country. This means that one country gets a PPA, i.e., trades at a virtual fixed price, while the other country pays the difference. In a deterministic setting it is possible to construct two PPAs, one for each country, such that the compensation to be paid to the first country according to its PPA exactly matches the compensation to be paid by the second country according to its PPA. This property is called \textit{budget balancedness} \cite{narahari2014game}. A multi-country PPA mechanism is proposed in \cite{kristiansen2018}. However, in a stochastic setting, a multi-country PPA mechanism cannot be constructed, as this would require different PPA prices for every scenario. We illustrate this in the following example.

\begin{example}
Consider a network consisting of two countries, A and B, with no initial transmission capacity. There is a proposed investment in a cable of capacity 10 between A and B. We consider two simplified, equally likely, one-period scenarios. In scenario 1 (indicated by the superscript 1), the flow through the new cable is $f_{AB}^1 = 10$, the prices in A and B are $\pi_A^1 = 1$, $\pi_B^1 = 2$, and the welfare effects of the new cable are $\Delta \text{TW}_A^1 = -10$ and $\Delta \text{TW}_B^1 = 20$. In scenario 2, we have $f_{AB}^2 = 10$, $\pi_A^2 = 1$, $\pi_B^2 = 3$, $\Delta \text{TW}_A^2 = -10$, and $\Delta \text{TW}_B^1 = 40$. Before realization of the uncertainty, we need to construct two PPAs that share the welfare gains/losses equally in expectation. Note that $\E[\Delta \text{TW}_A] = -10$ and $\E[\Delta \text{TW}_B] = +30$. Hence, on average, A should receive a compensation $C_A$ of 20 by B. For A, this yields the following equation for its PPA price:
\begin{align*}
    \E[C_A^{\text{PPA}_A}] = \E[f_{AB} (\pi^{\text{PPA}_A} - \pi_A)] = 20,
\end{align*}
which, after some calculation, yields $\pi^{\text{PPA}_A} = 3$. Similarly, for country B we have the equation
\begin{align*}
    \E[C_B^{\text{PPA}_B}] = \E[f_{BA} (\pi^{\text{PPA}_B} - \pi_A)] = -20,
\end{align*}
which yields $\pi^{\text{PPA}_B} = 4.5$. We observe that using these PPA prices, the compensation received by A equals the compensation paid by B, i.e., $\E[C_A^{\text{PPA}_A}] = -\E[C_B^{\text{PPA}_B}]$. So buget balancedness holds in expectation. 

Now consider a single scenario, e.g., scenario 1. For this scenario we have $C_A^{\text{PPA}_A,1} = f^1_{AB} (\pi^{\text{PPA}_A} - \pi^1_A) = 10 (3 - 1) = 20$. For B, we have $C_B^{\text{PPA}_B,1} = f^1_{BA} (\pi^{\text{PPA}_B} - \pi^1_B) = - 10 (4.5 - 2) = - 25$. Note that $C_A^{\text{PPA}_A,1} \neq - C_B^{\text{PPA}_B,1}$, so the compensations do not add to zero. Similarly, for scenario 2 we observe $C_A^{\text{PPA}_A,2} = 20$ and $C_B^{\text{PPA}_B,2} = -15$, which also do not add to zero. We conclude that in a stochastic setting, budget balancedness generally does not hold for PPAs. \hfill \qedsymbol
\end{example}

As a consequence of this lack of budget balancedness in a stochastic setting, we must pick one country that receives the PPA, i.e., that gets to trade at a virtual, fixed price. This is also the reason for the subscript $A$ in $C_i^{\text{PPA}_A}$, $i=A,B$, and $\pi^{\text{PPA}_A}$. Now, it is not hard to show that a Norway-based PPA yields larger compensations if $\pi^{\text{NO2}}$ is higher. Similarly, a Germany-based PPA yields larger compensations if $\pi^{\text{DE}}$ is higher. In Table~\ref{tab:price_correlations} we observe that in both these situations, Norway tends to profit more from the new line, while Germany suffers more. Hence, in situations where higher compensations are desired, both PPAs indeed yield higher compensations. This gives us some confidence that the PPAs might be able to succeed in mitigating risk for the countries involved. Based on the fact that the correlations with the NO2 price are stronger than those with the DE price, we also expect the Norway-based PPA to perform better than the Germany-based PPA.

\begin{table}[]
\centering
\caption{Correlations between welfare effect of the new transmission line in Norway and Germany and different price measures.}
\label{tab:price_correlations}
\begin{tabular}{llll}
\toprule
& \textbf{$\pi^{\text{NO2}}$} & \textbf{$\pi^{\text{DE}}$} \\ \midrule
\textbf{$\Delta \text{TW}_{\text{NO}}$} & 0.23 & 0.52 \\
\textbf{$\Delta \text{TW}_{\text{DE}}$} & -0.70 & -0.13 \\ \bottomrule
\end{tabular}%
\end{table}

\subsubsection{Novel mechanisms}
\label{subsubsec:novel_mechanisms}

Besides the lump sum and PPAs, we investigate the potential of other, novel compensation mechanisms. We aim for risk-mitigating mechanisms. Ideally, the mechanism is such that the compensation amount mimics the relative welfare effects in the countries.

One possibility would be to \textit{compute} the actual welfare effects using an economic model (such as the one presented in this paper) and base the compensation on this value. The advantage of such a model is that it is likely as close to compensating actual welfare effects as we can plausibly get, and hence, has the greatest potential for risk sharing. Specifically, we define the \textit{ideal} mechanism as the mechanism that directly shares the welfare benefits from the new transmission line among the participating countries in every scenario, according to some distribution rule represented by the coefficients $\lambda_i \geq 0$, $i \in I$, with $\sum_{i \in I} \lambda_i = 1$. That is, for every scenario $\omega \in \Omega$, the compensation to country $i$ in the set $I$ of participating countries is given by
\begin{align*}
    C^{\text{ideal},\omega}_i = \lambda_i \bigg(\sum_{j \in I} \Delta \text{TW}^\omega_j\bigg)  - \Delta \text{TW}^\omega_i.
\end{align*}
The coefficients $\lambda_i$, $i \in I$, should be chosen in such a way that in expectation, the total welfare gains are distributed according to the predetermined distribution rule (e.g., the Shapley value or an equal-share principle).

There are a number of downsides to this theoretically ideal mechanism, though. The main issue is that the compensation amounts depend on the welfare benefits to the countries \textit{as computed by the model}. Inevitably, however, the model will be imperfect and thus, the compensation levels are not ``correct''. Hence, conflicts may arise over the model to be used, which may undermine trust between the parties involved. Moreover, even if the model were perfect, it might be seen as a black box by non-experts; a simpler mechanism might be preferred. 

To construct other novel, risk-sharing compensation mechanisms, we take the following approach. We search for economic measures that (1) depend on the scenario, (2) relate to the new transmission line, and (3) are correlated with the welfare effects in the hosting countries. If we find such measures, then we can base a compensation mechanism on them.

Based on the first two conditions listed above, we propose two candidate measures: the amount of flow through the new line (referred to as \textit{flow}), and the economic value of the flow through the new line (referred to as \textit{flow value}). We define flow and flow value such that flow in the direction $\text{NO2}\to\text{DE}$ is counted as positive and flow in the opposite direction as negative. Moreover, note that the flow value is ambiguous: in periods in which the line is congested (i.e., flow equals transmission capacity), there is a price differential between the two connected nodes. In such cases, we compute the flow value by using the \textit{average} of the NO2 price and the DE price.

\begin{table}[]
\centering
\caption{Correlations between two flow-based measures and the welfare effect of the new transmission line in different countries.}
\label{tab:corr_country_measure}
\begin{tabular}{llllll}
\toprule
\textbf{}                 & \textbf{NO} & \textbf{AT} & \textbf{FR} & \textbf{DE} & \textbf{DK} \\ \midrule
\textbf{flow}             & 0.40        & -0.20       & -0.25       & 0.45        & 0.12        \\
\textbf{flow value} & 0.73        & 0.29        & 0.26        & -0.18       & -0.29      \\ \bottomrule
\end{tabular}%
\end{table}

\begin{figure}
    \centering
    \includegraphics[scale=0.4]{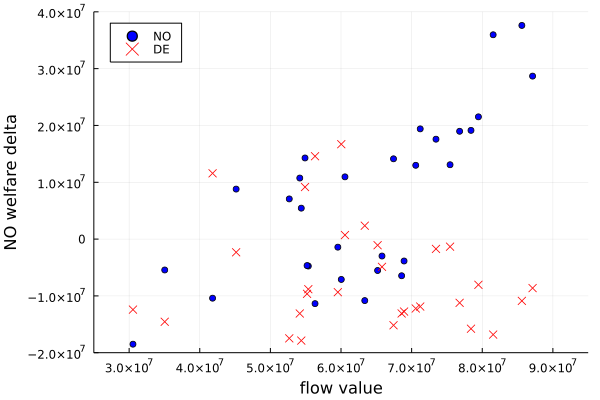}
    \caption{Scatter plot of the total value (using the hourly average of the NO2 and DE price) of the flow through the NO2-DE line and the welfare effect of the new line for Norway and Germany. Every dot/cross represents a scenario.}
    \label{fig:scatter_flow_value_vs_NO_DE_welfare_delta}
\end{figure}

To investigate the potential of these three candidate measures, we compute for each measure its correlations with the welfare effects of the new line in the hosting countries; see Table~\ref{tab:corr_country_measure}. Ideally, these correlations are strong and of opposite sign. We observe that flow value is positively correlated with the welfare effect of the new line in all three benefiting countries (Norway, Austria, and France), while negatively correlated with the welfare effect in the suffering countries (Germany and Denmark). Zooming in on the two hosting countries, we observe that it is particularly strongly correlated with the welfare effect in Norway. Moreover, in Figure~\ref{fig:scatter_flow_value_vs_NO_DE_welfare_delta} we observe that high values for the flow value correspond to scenarios in which Norway benefits most from the new line, while Germany tends to suffer. Based on these observations, flow value appears to be a good candidate measure to base a compensation mechanism on. 

For flow, on the other hand, the signs of the correlations do not line up with the sign of the expected welfare effect in each country. Moreover, the strongest correlation has a magnitude of 0.45, which is less than then 0.73 we observe for flow value. Hence, flow seems to be a weaker candidate measure to base a compensation mechanism on. Nevertheless, for the sake of completeness we will keep it in our list of candidates and test its performance rigorously.

Based on these two measures, we propose the following novel compensation mechanisms: the \textit{flow-based} compensation mechanism, under which country $i$ receives a compensation amount of
\begin{align*}
    C_i^{\text{flow},\omega} = \alpha_i \sum_{t \in \mathcal{T}} f_{\text{NO2-DE}}^{\omega,t},
\end{align*}
for some $\alpha_i \in \R$, and the \textit{value-based} compensation mechanism, under which country $i$ receives
\begin{align*}
    C_i^{\text{value},\omega} = \beta_i \sum_{t \in \mathcal{T}} f_{\text{NO2-DE}}^{\omega,t} \cdot \bar{\pi}^{\omega,t},
\end{align*}
for some $\beta_i \in \R$, where $\pi_t := \frac{1}{2} ( \pi^{NO2}_t + \pi^{DE}_t )$. The values of $\alpha_i$ and $\beta_i$, respectively, are chosen in such a way that on average, every country receives a compensation based on some predetermined rule (e.g., the Shapley value). In both cases, to achieve budget balancedness, the values of $\alpha$ and $\beta$, respectively, over all countries $I$ participating in the compensation scheme should add to zero, i.e., $\sum_{i \in I} \alpha_i = 0$ and $\sum_{i \in I} \beta_i = 0$.

\subsection{Performance}
\label{subsec:performance}
We now numerically investigate the performance of the various compensation mechanisms on the case study from Section~\ref{sec:TEP}. We implemented the following compensation mechanisms: no compensation, lump-sum compensation, a Norway-based PPA, a Germany-based PPA, a flow-based mechanism, a value-based mechanism, and a theoretically ideal model-based mechanism. In order to have a fair comparison, we parametrized each compensation mechanism in such a way that it yields the same expected compensation amount. The amount is such that the expected net welfare effect of the new line after compensation is equal in both countries. Incidentally, this coincides with the Shapley value proposed in \cite{kristiansen2018}. Since the expected compensation amount is equal among the mechanisms, a risk neutral country should be indifferent about them.

However, the performance of the mechanisms may vary in terms of the resulting risk that the countries are exposed to. We assume all countries are risk-averse and hence, we prefer mechanisms that reduce the amount of risk faced by each country as a consequence of the transmission investment. Note that the term ``risk'' is ambiguous; it is sometimes interpreted as analogous to ``variability'', and sometimes as ``likelihood and/or magnitude of losses'' \cite{rockafellar2007coherent}. In our analysis we don't choose one interpretation over the other, but include measures corresponding to both interpretations of risk.

\subsubsection{Bilateral compensation}

\begin{figure}
    \centering
    \includegraphics[scale=0.75]{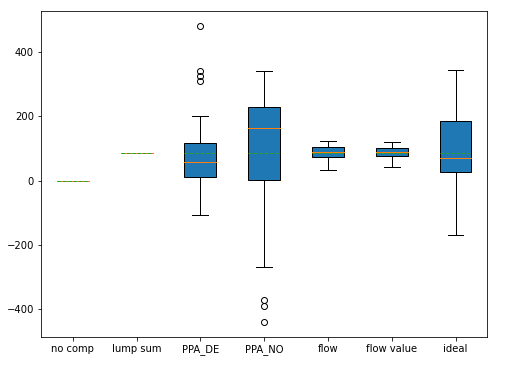}
    \caption{Boxplot of the compensation amount paid by Norway to Germany under various compensation mechanisms. The green dashed lines indicate averages, while the orange solid lines indicate medians.}
    \label{fig:boxplot_comp}
\end{figure}

First, we consider a setting with bilateral compensations between the two hosting countries Norway and Germany. We start by analyzing the compensation amount itself under various mechanisms. In Figure~\ref{fig:boxplot_comp} we present a boxplot of the compensation paid by Norway to Germany. Note that indeed, the lump sum leads to a fixed compensation amount, while the other mechanisms vary per scenario. The PPAs show a significantly lager variability than the flow- and value-based mechanisms. Hence, the latter two are more predictable, which may be seen as an advantage.

\begin{figure}[h]
    \centering
    \begin{subfigure}[b]{\columnwidth}
        \centering
        \includegraphics[scale=0.75]{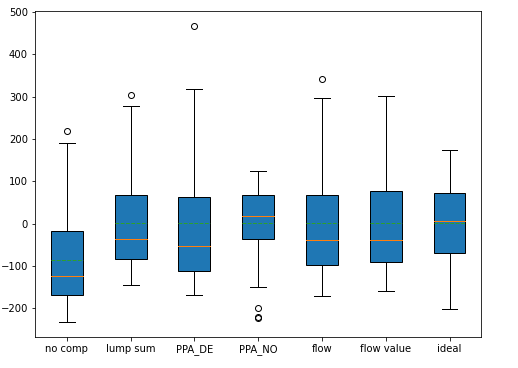}
        \caption{Germany}
        \label{subfig:boxplot_DE}
    \end{subfigure}\\
    \begin{subfigure}[b]{\columnwidth}
        \centering
        \includegraphics[scale=0.75]{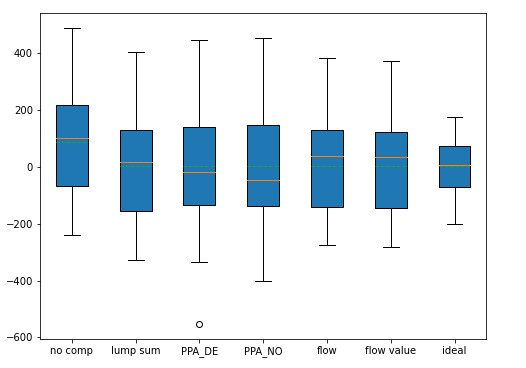}
        \caption{Norway}
        \label{subfig:boxplot_NO}
    \end{subfigure}
    \caption{Boxplots of the net welfare effects of the new line for Germany and Norway under various compensation mechanisms. The green dashed lines indicate averages, while the orange solid lines indicate medians.}
    \label{fig:boxplot_DE_NO}
\end{figure}

More important, however, is how the different mechanisms translate into net welfare effects of the new line. In particular, we are interested in the corresponding risk faced by each country. We first focus on variability-based measures of risk. For this purpose, we refer to the boxplots in Figure~\ref{fig:boxplot_DE_NO} and the first two columns of Table~\ref{tab:compensation_statistics}. Note that the theoretical ideal mechanism is able to significantly reduce these measures of variability, though not completely. This is due to the fact that there is variability in the combined welfare effects of the new line, aggregated over both countries. Hence, we should not expect to be able to eliminate all variability using compensation mechanisms, although the ideal mechanism shows that significant reduction is theoretically possible.

Next, we observe that a Germany-based PPA increases the risk faced by both countries, even that of Germany. Hence, the variability in the the compensation amount observed in Figure~\ref{fig:boxplot_comp} does not cancel out the variability in the welfare effects before compensation. This may be caused by the fact that the correlations between the welfare effects of the new line in Norway and Germany and the price in Germany are not large enough. 

A Norway-based PPA, however, reduces the risk faced by Germany significantly, to a level close to the theoretical ideal mechanism, while increasing that faced by Norway. Here, at least for Germany, the correlation between the welfare effect in Germany and the NO2 price is apparently large enough for the variability in the compensation amount to cancel out variability in the welfare effect of the new cable in Germany. For Norway, however, the corresponding effect is not strong enough, which might be explained by the weaker correlation between the NO2 price and the welfare effect of the new line on Norway, observed in Table~\ref{tab:price_correlations}. Hence, the net effect of the Norway-based PPA is that it shifts risk from Germany to Norway. This may or may not be desirable, depending on the risk preferences of the two countries. 

Finally, we observe that the value-based mechanism outperforms all other measures for both countries, except for the Norway-based PPA, which is the best option for Germany. Notably, the value-based mechanism succeeds in reducing the amount of risk faced by both Norway and Germany, compared to a lump sum. Hence, it succeeds in its purpose of risk mitigation. However, the improvement over the lump sum is only modest. Comparing the value-based mechanism to the theoretical ideal, we see that there is still a significant potential for improvement.

\begin{table*}[t]
\centering
\caption{Various measures capturing the level of risk faced by Germany and Norway as a result of the new line, after compensation by different mechanisms. The measures are: standard deviation of the compensation amount, standard deviation of the net total welfare effect, probability of welfare loss, expected welfare loss, CVaR of welfare loss (with parameter 0.80). All numbers (except for the percentages) are in millions of euros annually.}
\label{tab:compensation_statistics}
\begin{tabular}{l|l|ll|ll|ll|ll}
\toprule
mechanism & \textbf{std(C)} & \multicolumn{2}{c|}{\textbf{std($\Delta \text{NTW}$)}} & \multicolumn{2}{c|}{\textbf{P(L)}} & \multicolumn{2}{c|}{\textbf{E{[}L{]}}} & \multicolumn{2}{c}{$\textbf{CVaR}_{\textbf{80}}\textbf{(L)}$} \\ 
    &       & DE    & NO    & DE   & NO   & DE    & NO   & DE    & NO    \\ \midrule
no comp.   & 0.0   & 124.8 & 188.8 & 80\% & 43\% & 109.1 & 40.5 & 212.1 & 140.4 \\
lump sum   & 0.0   & 124.8 & 188.8 & 63\% & 47\% & 49.2  & 78.6 & 125.4 & 227.1 \\
PPA DE     & 130.5 & 156.4 & 212.1 & 57\% & 53\% & 58.9  & 82.5 & 145.2 & 275.5 \\
PPA NO     & 211.2 & 94.9  & 209.7 & 37\% & 53\% & 34.8  & 87.2 & 150.8 & 272.0 \\
flow       & 22.4  & 136.4 & 181.1 & 57\% & 43\% & 52.4  & 76.0 & 144.8 & 230.8 \\
flow value & 19.3  & 122.8 & 175.2 & 57\% & 47\% & 48.1  & 73.9 & 131.2 & 213.8 \\
ideal	   & 130   & 93.6  & 93.6  & 47\% & 47\% & 38.0  & 38.0 & 124.8 & 124.8 \\
\bottomrule
\end{tabular}%
\end{table*}

Next, in the remaining columns of Table~\ref{tab:compensation_statistics} we consider loss-oriented measures. We compute the probability of a net welfare loss, the expected net welfare loss (only scenarios with losses contribute to this value; gains are regarded as zeros), and the $80\%$ conditional value at risk (CVaR) \cite{rockafellar2002conditional} of the net welfare loss, representing the expected value of the $20\%$ worst cases. We observe that the theoretical ideal achieves in mitigating risk significantly, although not completely. The fact that it cannot completely eliminate the risk of loss was already argued in Section~\ref{subsec:mechanisms} and illustrated by Figure~\ref{fig:scatter_NO_DE_welfare_delta}. However, Table~\ref{tab:compensation_statistics} shows that significant improvements are possible, at least in theory. 

Turning to the other mechansims, we observe that the Germany-based PPA performs bad on all fronts. For the Norway-based PPA we see a similar performance as before (good for Germany, bad for Norway), except according to the CVaR measure: in terms of CVaR it performs relatively bad for Germany as well. The interpretation is that although the expected losses to Germany are small, the very worst scenarios are worse than under other mechanisms. Next, the value-based mechanism outperforms all other mechanisms (except the theoretical ideal) according to most measures. It outperforms the flow-based mechanism on all measures except the probability of a loss in Norway. Moreover, it outperforms the lump sum on all but one measure: CVaR for Germany. Hence, again, it seems to succeed in its purpose of mitigating risk for most countries. However, comparing it with the theoretical ideal mechanism, there is again significant room for improvement.

We can draw the following conclusions. Firstly, the theoretical ideal mechanism shows that there is indeed a significant potential for mitigating risk by using scenario-dependent compensations. Secondly, a Germany-based PPA performs by far the worst of all measures considered. It increases the risk faced by the countries compared to a lump sum. However, a Norway-based PPA does yield good results for Germany. It basically transfers part of the risk from Germany to Norway. Such behavior might or might not be desirable, depending on the relative risk preferences of both countries. Next, the value-based mechanism consistently outperforms the flow-based mechanism. Hence, including the additional price information allows the mechanism to succeed better in risk sharing. 

Finally, the value-based mechanism outperforms most other mechanisms according to most measures. In particular, it outperforms the lump sum on all measures except CVaR for Germany. Hence, it seems to succeed in achieving what it was constructed to do: reducing the risk faced by both hosting countries. However, the value-based mechanism only improves upon a lump sum by a modest degree. Comparing this with the theoretical ideal mechanism, which performs much better, we conclude that there is still significant room for improvement. In particular, a mechanism based on some measure exhibiting stronger correlations with the welfare effects in Norway and Germany than flow value does in Table~\ref{tab:corr_country_measure}, might achieve higher levels of risk mitigation.

\subsubsection{Multilateral compensation}

In practice, countries hosting a planned transmission expansion might want to involve other countries that are affected by the planned investment. One reason for this may be to avoid conflict that might hurt future cooperation. In our case study in particular, there are three other countries that are significantly affected by the proposed transmission expansion: Austria, France, and Denmark. Even though the welfare gain for Norway is sufficient to compensate Germany on its own, the majority of the welfare gains actually falls on other countries, most notably Austria and France. Moreover, besides Germany, Denmark is another high-production country that is hurt by the cheap Norwegian power entering the European market. Hence, involving all these five countries in a compensation scheme might be a desirable course of action.

Given this motivation, we now investigate the performance of different compensation mechanisms in a five-country coalition consisting of the countries mentioned above. We compare lump sum payments to the flow- and value-based mechanisms proposed in Section~\ref{subsubsec:novel_mechanisms}. Note that the definition of PPAs is unclear in this multilateral setting. For instance, suppose we give Norway a PPA, then it is unclear how we should use this to determine the compensation paid from Austria to Denmark. For this reason, we disregard PPAs in this analysis. Finally, in practice, the share of the benefits allocated to each country is a topic for negotiations. In this case study, we assume an equal-share principle, which entails that every mechanism is parametrized such that the expected net total welfare effect of the new transmission line after compensation is equal in all five countries (25.6 million euros annually).

\begin{table*}[t]
\centering
\caption{Various measures capturing the level of risk faced by each coalition country as a result of the new line, after compensation by different mechanisms. All numbers (except for the percentages) are in millions of euros annually.}
\label{tab:compensation_statistics_coalition}
\begin{tabular}{cllllllll}
\toprule
\multicolumn{1}{l}{\textbf{}} &
  \textbf{mechanism} &
  \textbf{std(C)} &
  \textbf{std($\Delta\text{NTW}$)} &
  \textbf{P(L)} &
  \textbf{E{[}L{]}} &
  \textbf{$\text{CVaR}_{\text{0.8}}\text{(L)}$} \\ \midrule
\multirow{4}{*}{NO} & no comp.   & 0.0  & 188.8 & 43\% & 40.5  & 140.4 \\
                    & lump sum   & 0.0  & 188.8 & 43\% & 67.7  & 203.1 \\
                    & flow       & 16.2 & 183.0 & 43\% & 66.2  & 205.8 \\
                    & flow value & 14.0 & 178.9 & 43\% & 64.5  & 193.5 \\ 
                    & ideal      & 159.8& 43.2  & 27\% & 7.0   & 33.0  \\ \midrule
\multirow{4}{*}{AT} & no comp.   & 0.0  & 79.7  & 0\%  & 0.0   & 0.0   \\
                    & lump sum   & 0.0  & 79.7  & 40\% & 15.5  & 61.4  \\
                    & flow       & 23.0 & 87.2  & 47\% & 20.1  & 72.4  \\
                    & flow value & 19.9 & 76.3  & 43\% & 15.3  & 50.3  \\
                    & ideal      & 63.5 & 43.2  & 27\% & 7.0   & 33.0  \\ \midrule
\multirow{4}{*}{FR} & no comp.   & 0.0  & 54.2  & 0\%  & 0.0   & 0.0   \\
                    & lump sum   & 0.0  & 54.2  & 30\% & 8.0   & 38.2  \\
                    & flow       & 11.5 & 58.1  & 37\% & 9.7   & 42.2  \\
                    & flow value & 9.9  & 52.5  & 37\% & 7.3   & 31.5  \\ 
                    & ideal      & 47.8 & 43.2  & 27\% & 7.0   & 33.0  \\ \midrule
\multirow{4}{*}{DE} & no comp.   & 0.0  & 124.8 & 80\% & 109.1 & 212.1 \\
                    & lump sum   & 0.0  & 124.8 & 60\% & 34.2  & 101.4 \\
                    & flow       & 28.6 & 140.0 & 53\% & 40.3  & 128.2 \\
                    & flow value & 24.7 & 122.8 & 53\% & 34.9  & 110.9 \\ 
                    & ideal      & 125.4& 43.2  & 27\% & 7.0   & 33.0  \\ \midrule
\multirow{4}{*}{DK} & no comp.   & 0.0  & 36.1  & 97\% & 60.2  & 112.7 \\
                    & lump sum   & 0.0  & 36.1  & 20\% & 5.4   & 27.0  \\
                    & flow       & 22.1 & 44.6  & 30\% & 9.0   & 39.4  \\
                    & flow value & 19.1 & 35.7  & 27\% & 5.4   & 25.5  \\
                    & ideal      & 70.1 & 43.2  & 27\% & 7.0   & 33.0  \\ \bottomrule
\end{tabular}%
\end{table*}

In Table~\ref{tab:compensation_statistics_coalition} we present performance measures for the various compensation mechanisms. First, we observe that under all mechanisms (except the theoretical ideal), Norway and Germany face by far the largest levels of risk. This can be understood by the fact that since they host the new cable, the cable affects these countries most directly. Next, we observe that the value-based mechanism outperforms the flow-based mechanism on almost all measures for all countries. This is in line with our results in the bilateral case and can again be explained by the more desirable correlations observed in Table~\ref{tab:corr_country_measure} 

The question remains which of the remaining two mechanisms, lump sum and the value-based mechanism, is preferable. In terms of a variability-based definition of risk, the value-based mechanism is clearly the winner. It has a smaller standard deviation of net total welfare effects than the lump sum for all countries. Hence, it indeed succeeds in risk mitigation. Next, in terms of a loss-based definition of risk, the results are again in favor of the flow-value based mechanism, though less strongly so. In terms of expected loss and CVaR of loss, all countries except for Germany are better off with the value-based mechanism. The reason for underperformance for Germany may be the weak correlation between flow value and the welfare effect of the new line in Germany; see Table~\ref{tab:corr_country_measure}.

Overall, we conclude that the value-based mechanism consistently outperforms the flow-based mechanism and also outperforms the lump sum in terms of most measures for most countries. All in all, the value-based mechanism seems the most promising in terms of mitigating risk of the countries involved. However, comparing it to the theoretical ideal mechanism, there is again significant room for improvement.

\section{Conclusion}
\label{sec:conclusion}

We investigated the potential of existing and novel welfare compensation mechanisms in TEP under uncertainty. The simplest existing mechanism, lump sum payments, does not take uncertainty into account at all. Other mechanisms, under which the compensation amount depends on the scenario, can potentially exploit negative correlations between the welfare effects of a new transmission line in benefiting and suffering countries in order to mitigate the risk of all parties involved. 

We conducted numerical experiments in a case study of the Northern-European power sector. We find a system-optimal investment in a new transmission line between Norway and Germany that benefits Norway but hurts Germany in terms of expected total welfare. We considered both bilateral compensations (between Norway and Germany alone) and multilateral compensations (involving other affected countries, too). In both settings, we observed that a theoretically ideal, model-based mechanism can significantly reduce the levels of risk faced by the countries involved. This highlights the relevance of our research and the potential of mitigating risk by using scenario-dependent compensation mechanisms.

We analyzed one scenario-dependent mechanism from the literature: PPAs. We first show that in a stochastic setting, budget balancedness does not hold for PPAs. Hence, one should select a single country that receives the PPA. Moreover, in the numerical experiments we observed that a Germany-based PPA \textit{increases} risk for both countries, while a Norway-based PPA shifts risk from Germany to Norway. We conclude that when considering a PPA, one should be careful in selecting the country at which the PPA is based.

We also tested two novel mechanisms, based on the flow through the new line and its economic value. In both the bilateral and multilateral setting, our novel value-based mechanisms performs best in terms of mitigating risk for the countries involved. It appears to do so by successfully exploiting negative correlations in the welfare effects of the new transmission line between benefiting and suffering countries. In particular, the value-based mechanism outperforms the lump sum payment. However, the improvement is only moderate. Comparing it with the theoretically ideal mechanism, we see that there is still significant potential for improvement. We expect that the level of outperformance may be higher if the negative correlations mentioned above are stronger or if the correlations between the economic value of flow through the new transmission line and the welfare effects of this line in the neighboring countries are stronger. 
 
While we deem the value-based mechanism most promising, its performance may well depend on the specific practical setting at hand. Therefore, in practical situations, we suggest to run an analysis like the one presented in this paper before choosing a particular compensation mechanism. Note that if the proposed transmission expansion plan is found by running a TEP model, then the proposed analysis can be performed by a simple extension of this model. One might use this paper as a blueprint for such an analysis. In any case, we suggest to include our value-based mechanism as one of the candidate compensation mechanisms.

Future research might focus on further investigating the performance of various compensation mechanisms in settings beyond the case study investigated in our paper. In particular, it would be interesting to test our hypothesis that the value-based mechanism performs better in situations with a strong correlation between the economic value of the flow through the new transmission line and its welfare effects in the neighboring countries. In addition, novel mechanisms may be developed based on measures exhibiting such strong correlations, or mixtures of several mechanisms may be tested. Moreover, it would be interesting to investigate the performance of different compensation mechanisms in a setting with investments in multiple transmission lines simultaneously, rather than the single-line setting used in this paper.

Another avenue for future research might be to investigate the incentive-distorting effects of welfare compensations and various mechanisms in particular, as sketched in Section~\ref{subsec:incentive_effects}. It would be interesting to see how much welfare compensations may cause the system to deviate from the equilibrium and what the welfare effects of these deviations are. Such an analysis might be able to identify the types of government interventions (taxes/subsidies) and compensation mechanisms that minimize this issue of incentive distortions.

\paragraph{Acknowledgements}
We want to thank THEMA Consulting for the data they shared with us.

\begin{appendices}

\section{Mathematical model}
\label{sec:mathematical_model}

Our TEP model consists of two levels. In the lower level, the producers, consumers and the TSO act in the electricity market. We assume that each of these actors maximizes their own surplus and we assume perfect competition with all actors being price takers. The equilibrium problem arising from these interacting optimization problems can be formulated as a mixed-complementarity program (MCP) consisting of the Karush-Kuhn-Tucker (KKT) conditions of each actor's optimization problem, combined with a market clearing condition. In the upper level, a social planner is endowed with the task of deciding the transmission expansion investment levels. We assume that the social planner acts as a Stackelberg leader that tries to maximize total welfare of the entire system, while taking the optimal decisions of the followers into account. Together, the bi-level model constitutes an MPEC.

\subsection{Notation}
\noindent\textbf{Sets:}

\renewcommand{\arraystretch}{1.5}
\begin{tabularx}{0.95\linewidth}{@{}>{\bfseries}l@{\hspace{.5em}}X@{}}
$\mathcal{N}$ & Set of nodes (indexed by $n$) \\
$\mathcal{L}$ & Set of lines (indexed by $l$) \\
$\mathcal{G}_n$ & Set of generators in node $n$ (indexed by $g$) \\
$\mathcal{R}_n$ & Set of renewables in node $n$ (indexed by $r$) \\
$\mathcal{S}$ & Set of seasons (indexed by $s$) \\
$\mathcal{T}$ & Set of time periods (indexed by $t$) \\
$\mathcal{T}_S$ & Set of time periods in season $s$ (indexed by $t$) \\
$\Omega$ & Set of scenarios (indexed by $\omega$) \\
\end{tabularx}
\\

\noindent\textbf{Parameters:}

\renewcommand{\arraystretch}{1.5}
\begin{tabularx}{0.95\linewidth}{@{}>{\bfseries}l@{\hspace{.5em}}X@{}}
$P_\omega$ & Probability for scenario $\omega$ \\
$I^{R}_{\omega rt}$ & Production profile for renewable $r$ in scenario $\omega$ in time period $t$ \\
$G^{R}_{r}$ & Installed capacity for renewable $r$ $[\SI{}{\mega\watt}]$\\
$C^{I,R}_{r}$ & Investment cost for renewable $r$ [$\SI{}{\text{\euro}\per\mega\watt}$]\\
$C^{I,G}_{g}$ & Investment cost for generator $g$ [$\SI{}{\text{\euro}\per\mega\watt}$]\\
$C^{G}_{gt}$ & Marginal cost for generator $g$ in time period $t$ [$\SI{}{\text{\euro}\per\mega\watt}$]\\
$G^{Max}_{g}$ & Installed generation capacity for generator $g$ [$\SI{}{\mega\watt}$]\\
$Q^{Max}_{\omega gs}$ & Production limit for generator $g$ in scenario $\omega$ in season $s$ [$\SI{}{\mega\watt h}$]\\
$A_{nl}$ & Node-line incidence matrix entry for node $n$ and line $l$\\
$C^{I,L}_{l}$ & Investment cost for line $l$ [$\SI{}{\text{\euro}\per\mega\watt}$]\\
\end{tabularx}
\renewcommand{\arraystretch}{1.5}
\begin{tabularx}{0.95\linewidth}{@{}>{\bfseries}l@{\hspace{.5em}}X@{}}
$F^{Max}_{l}$ & Maximum line capacity for line $l$ [$\SI{}{\mega\watt}$]\\
$D^{A}_{\omega nt}$ & Slope of inverse demand function for node $n$ in scenario $\omega$ in time period $t$\\
$D^{B}_{\omega nt}$ & Constant of inverse demand function for node $n$ in scenario $\omega$ in time period $t$\\
\end{tabularx}
\\

\noindent\textbf{Variables:}

\begin{tabularx}{0.95\linewidth}{@{}>{\bfseries}l@{\hspace{.5em}}X@{}}
$y^R_{r}$ & Capacity expansion for renewable $r$ [$\SI{}{\mega\watt}$] \\
$y_{g}$ & Capacity expansion for generator $g$ [$\SI{}{\mega\watt}$] \\
$q_{\omega gt}$ & Production for generator $g$ in scenario $\omega$ in time period $t$ [$\SI{}{\mega\watt}$] \\
$f_{\omega lt}$ & Flow in line $l$ in scenario $\omega$ in time period $t$ [$\SI{}{\mega\watt}$] \\
$x_{l}$ & Capacity expansion for line $l$ [$\SI{}{\mega\watt}$] \\
$d_{\omega nt}$ & Demand in node $n$ in scenario $\omega$ in time period $t$ [$\SI{}{\mega\watt}$] \\
$\pi_{\omega nt}$ & Price in node $n$ in scenario $\omega$ in time period $t$ [$\SI{}{\text{\euro}\per\mega\watt}$] 
\end{tabularx}

\subsection{Lower-level problem}
\label{subsec:lower-level_problem}
In this subsection we describe the lower-level equilibrium problem. This equilibrium problem can be represented as an MCP consisting of the KKT conditions of the optimization problems of all the actors in the market, combined with a market clearing constraint. We will not present the KKT conditions explicitly, but simply state each actor's optimization problem.

\subsubsection{Conventional energy producer problem}
\label{subsubsec:producer_problem}
Every conventional generating unit (power plant) is modeled as an independent generating company maximizing its operational profits minus investment costs. Each generating company can freely choose its generation levels to maximize their profits. Moreover, it has the possibility to invest in additional generating capacity if needed. Eq.~\eqref{opt:GenCoProblem} describes the optimization problem for generator $g \in \mathcal{G}_n$ located in node $n \in \mathcal{N}$, where the constraints are defined $\forall \: \omega \in \Omega, s \in \mathcal{S}, t \in \mathcal{T}$.

\begin{maxi!}[1]
{q_{\omega gt}, y_{g}}
{\sum_{\omega \in \Omega}\sum_{t \in \mathcal{T}}P_\omega\left(\pi_{\omega nt}-C^G_{gt}\right)q_{\omega gt}-C^{I,G}_{g}y_g\label{opt:GenCoObjective}}
{\label{opt:GenCoProblem}}
{}
\addConstraint{q_{\omega gt}}{\leq G^{Max}_{g}+y_{g}\label{opt:GenCoProdCap}}
\addConstraint{\sum_{t\in\mathcal{T}_S}q_{\omega gt}}{\leq Q^{Max}_{\omega gs}\label{opt:GenCoEnergyLimit}}
\addConstraint{q_{\omega gt}, y_{g}}{\geq 0\label{opt:GenCoLargerThanZero}}
\end{maxi!}
The objective function in Eq.~\eqref{opt:GenCoObjective} consists of maximizing the expected revenue minus the investment costs in new generating capacity. Operating costs and investment costs are both assumed to be linear. Eq.~\eqref{opt:GenCoProdCap} states that the production must be no more than the existing generating capacity plus the invested generating capacity. Eq.~\eqref{opt:GenCoEnergyLimit} states that the total production in a season must be no more than the available quantity, which may vary per season. Finally, Eq.~\eqref{opt:GenCoLargerThanZero} states that the generation and investment quantities must be non-negative.

\subsubsection{Renewable energy producer problem}
\label{subsubsec:RES_problem}
The renewable energy companies want maximize their profit earned from generating and selling renewable energy. In contrast with the conventional power producers, they are not able to freely choose their production levels; these are determined by the wind and solar profile. They can choose to invest in additional generating capacity, though. The optimization problem for renewable energy producer $r \in \mathcal{R}_n$ located in node $n \in \mathcal{N}$ is given by Eq.~\eqref{opt:RESProblem}.
\begin{maxi!}[1]
{y^R_{r}}
{\sum_{\omega\in\Omega}\sum_{t\in\mathcal{T}}P_\omega\pi_{\omega nt}\left(G^R_{r}+y^R_{r} \right) I^R_{\omega rt}-C^{I,R}_r y^R_r\label{opt:RESObjective}}
{\label{opt:RESProblem}}
{}
\addConstraint{y_{r}^R}{\geq 0\label{opt:RESConNonneg}}
\end{maxi!}
The objective function in Eq.~\eqref{opt:RESObjective} consists of maximizing the expected revenue minus the investments in new renewable capacity. Investment costs are assumed to be linear. Eq.~\eqref{opt:RESConNonneg} states that capacity investment is non-negative.

\subsubsection{Consumer problem}
\label{subsubsec:consumer_problem}
The consumers aim to satisfy their demand for power at the lowest possible cost. In other words, they want to maximize the consumer surplus in each node $n$. Assuming a linear demand function, Eq.~\eqref{opt:ConsumerProblem} represents the maximization problem for the consumers located in node $n$.

\begin{maxi!}[1]
{d_{\omega nt}}
{\sum_{\omega\in\Omega}\sum_{t\in\mathcal{T}}P_\omega\left(\frac{1}{2}D^A_{\omega nt}d_{\omega nt}+D^B_{\omega nt}-\pi_{\omega nt}\right)d_{\omega nt}\label{opt:ConsumerObjective}}
{\label{opt:ConsumerProblem}}
{}
\end{maxi!}
Eq.~\eqref{opt:ConsumerObjective} is the objective function for the consumers, representing the consumer surplus.

\subsubsection{TSO Problem}
\label{subsubsec:TSO_problem}
We assume a single TSO that operates all lines. The TSO maximizes the expected congestion rent for all lines, and sets the line flows accordingly. Recall that we assume that the TSO is a price taker. Hence, it will not use its market power and it will basically act as a dummy player (this is confirmed by our finding that in the equilibrium solution the TSO makes zero profit in expectation). The optimization problem for the TSO is given by Eqs.~\eqref{opt:TSOProblem}, where the constraints are defined $\forall \: \omega \in \Omega, l \in \mathcal{L}, t \in \mathcal{T}$.

\begin{maxi!}[1]
{f_{\omega lt}, x_{l}}
{-\sum_{\omega \in \Omega}\sum_{i \in \mathcal{N}}\sum_{l\in \mathcal{L}}\sum_{t \in \mathcal{T}}P_{\omega}A_{nl}f_{\omega lt}\pi_{\omega nt}\label{opt:TSOObjective}}
{\label{opt:TSOProblem}}
{}
\addConstraint{f_{\omega lt}}{\leq F^{Max}_{l}+x_{l}\label{opt:TSOflowPos}}
\addConstraint{f_{\omega lt}}{\geq -F^{Max}_{l}-x_{l}\label{opt:TSOflowNeg}}
\end{maxi!}
The objective function in Eq.~\eqref{opt:TSOObjective} consists of the expected congestion rent earned from all lines. Eqs.~\eqref{opt:TSOflowPos} and \eqref{opt:TSOflowNeg} state that the flow in a line must not exceed the current capacity plus the invested capacity. Note that the optimization model can be equivalently separated into optmization problems for each line. Hence, the assumption of a single TSO is without loss of generality.

\subsubsection{Market clearing}
The market clearing constraint is used to connect the market actors' decisions together. It guarantees that the market clears, i.e., that supply meets demand. It is given by

\begin{equation}
 d_{\omega nt}+\sum_{l\in \mathcal{L}}A_{nl}f_{\omega lt}=\sum_{g\in \mathcal{G}_n}q_{\omega gt}+\sum_{r\in \mathcal{R}_n}\left(G^R_{r}+y^R_{r}\right)I^R_{\omega rt}  \label{opt:MarketClearing}  
\end{equation}
In particular, Eq.~\eqref{opt:MarketClearing} states that for a given node, the sum of demand and net outgoing flows must be equal to the total amount of power generated from both conventional and renewable sources. The market price is given by the dual variable $\pi_{\omega nt}$ corresponding to this constraint.

\subsection{Upper-level problem} \label{subsec:upper-level_problem}
The upper-level problem consists of the social planner's transmission expansion problem. The social planner chooses the investment levels $x_l$, $l \in \mathcal{L}$, that maximize the expected net total welfare in the system. Here, we define net total welfare as gross total welfare (consisting of producer surplus, consumer surplus, and congestion rent) minus investment cost (which are assumed to be linear). The social planner acts as a Stackelberg leader that takes the other actors' optimal responses into account. 

Let $TW(x,q,y,y^R,d,f,\pi)$ denote the expected gross total welfare corresponding to the decision vectors $x,q,y,y^R,d,f$ and price vector $\pi$. That is, $TW$ is equal to the sum of the objective functions of the optimization problems of all producers, consumers, and the TSO. Let $KKT_1, \ldots, KKT_4$ denote the sets of KKT conditions corresponding to the problems in Eqs.~\eqref{opt:GenCoProblem}-\eqref{opt:TSOProblem}, respectively. Then, the social planner's problem can be described as an MPEC of the form

\begin{maxi!}[1]
{x_{l} \geq 0}
{TW(x,q,y,y^R,d,f,\pi) - \sum_{l\in\mathcal{L}}C^{I,L}_l x_l}
{\label{opt:GovProblem}}
{\label{opt:GovObjective}}
\addConstraint{KKT_1, \ldots, KKT_4}{\label{opt:GovKKT}}
\addConstraint{\text{Market clearing condition Eq. } \eqref{opt:MarketClearing}}{\label{opt:GovMarketClearing}}
\end{maxi!}
Eq.~\eqref{opt:GovObjective} represents the social planner's objective, consisting of maximizing the net total welfare. Eqs.~\eqref{opt:GovKKT}-\eqref{opt:GovMarketClearing} are the equilibrium constraints arising from the lower-level MCP.

\subsubsection{Quadratic programming reformulation}
The social planner's optimization problem, given by the MPEC Eq.~\eqref{opt:GovProblem}, can be reformulated as a single quadratic program. To show this, first define $TW^*(x)$ as the value of $TW(x,q,y,y^R,d,f,\pi)$ at an optimal solution to the lower-level MCP defined by Eqs.~\eqref{opt:GovKKT}-\eqref{opt:GovMarketClearing} (this value is unique, as we will prove shortly). Then, the social planner's problem consists of maximizing net total welfare:

\begin{maxi!}[1]
{x_{l} \geq 0}
{TW^*(x) - \sum_{l\in\mathcal{L}}C^{I,L}_l x_l}
{\label{opt:GovProblemReformulated}}
{}
\end{maxi!}

Similar to the classical result by \cite{samuelson1952}, it can be shown that the lower-level MCP is equivalent to a central planner quadratic optimization problem in which total welfare is maximized. The proof of this equivalence, which we omit for brevity, is through the observation that the KKT conditions to the quadratic program, which are necessary and sufficient, are equivalent to the mixed-complementarity Eqs.~\eqref{opt:GovKKT}-\eqref{opt:GovMarketClearing} defining the lower-level MCP. Observe that $TW^*(x)$ denotes the optimal value of the quadratic program, which proves that this value is indeed unique. 

Substituting the lower-level quadratic programming formulation for $TW^*(x)$ into Eq.~\eqref{opt:GovProblemReformulated}, and combining the upper and lower level decisions into a single optimization problem yields a quadratic programming reformulation for the social planner's MPEC. It is given by Eq.~\eqref{opt:CentralPlanner}, where the constraints are defined $\forall \: \omega \in \Omega, g \in \mathcal{G}_n, r \in \mathcal{R}_n, n \in \mathcal{N}, l \in \mathcal{L},  s \in \mathcal{S}, t \in \mathcal{T}$.

\begin{maxi!}[1]
{\substack{q_{\omega gt}, y_{g} \\ y^R_{r}  d_{\omega nt} \\ f_{\omega lt}, x_{l}}}
{\sum_{\omega\in\Omega}\sum_{i\in\mathcal{N}}\sum_{t\in\mathcal{T}}P_\omega\left(\frac{1}{2}D^A_{\omega nt}d_{\omega nt}+D^B_{\omega nt}\right)d_{\omega nt}}{\label{opt:CentralPlanner}}{} \nonumber
\breakObjective{-\sum_{\omega\in\Omega}\sum_{g\in\mathcal{G}_n}\sum_{t\in\mathcal{T}}P_\omega C^G_{gt}q_{\omega gt} -\sum_{g\in\mathcal{G}_n}C^{I,G}_{g}y_g} \nonumber
\breakObjective{-\sum_{r\in\mathcal{R}_n}C^{I,R}_ry^R_r-\sum_{l\in\mathcal{L}}C^{I,L}_lx_l \label{opt:CentralPlannerObj3}} 
\addConstraint{q_{\omega gt}}{\leq G^{Max}_{g}+y_{g} \label{opt:CentralPlannerConFirst}}
\addConstraint{\sum_{t\in\mathcal{T}_S}q_{\omega gt}}{\leq Q^{Max}_{\omega gs}}
\addConstraint{d_{\omega nt}+\sum_{l\in \mathcal{L}}A_{nl}f_{\omega lt}}{=\sum_{g\in \mathcal{G}_n}q_{\omega gt}+\sum_{r\in \mathcal{R}_n}\left(G^R_{r}+y^R_{r}\right)I^R_{\omega rt}}
\addConstraint{f_{\omega lt}}{\leq F^{Max}_{l}+x_{l}}
\addConstraint{f_{\omega lt}}{\geq -F^{Max}_{l}-x_{l}}
\addConstraint{q_{\omega gt}, d_{\omega nt}}{\geq 0}
\addConstraint{y_{g}, y^R_{r}, x_{l}}{\geq 0 \label{opt:CentralPlannerConLast}}
\end{maxi!}
Here, the objective function in Eq.~\eqref{opt:CentralPlannerObj3} consists of the sum of the objective functions of all market participants' optimization problems less investment cost in new transmission lines. The constraints in Eqs.~\eqref{opt:CentralPlannerConFirst}--\eqref{opt:CentralPlannerConLast} are a concatenation of the constraints from all market actors' individual optimization problems and the market clearing constraint.

\section{Data}
\label{sec:data}
The model presented in Section~\ref{sec:mathematical_model} is solved using a power system consisting of Norway, Germany and their neighbouring regions. In this section we describe the data used to parametrize the model.

Marginal investment costs of generators and transmission lines are estimated using data on historical investments 
\cite{Blumsac2022Basic,schroder2013current,NorNed2008longest,Statnett2013Cooperation,Powerengineeringinternational2012Germany,Northconnect2022FAQ} and annualized using economic lifetimes based on \cite{schroder2013current} using an assumed interest rate of 4\%. The estimates are given in  Table~\ref{tab:InvestmentCosts in sec:Results}. Since our main concern is the interaction between Norway and Germany, all capacity expansions in transmission and generation in and between other countries than Norway and Germany are fixed at zero.

\begin{table}[t] 
\centering
\caption{Marginal annualized investment costs [\SI{}{\text{\EUR{}}\per\mega\watt}] for generating technologies and transmission lines.}
\label{tab:InvestmentCosts in sec:Results}
\begin{tabular}{@{}lrllr@{}}
\toprule
\multicolumn{2}{c}{\textbf{Generators}} & \multicolumn{1}{c}{\textbf{}} & \multicolumn{2}{c}{\textbf{Lines}} \\ \midrule
Coal                & 136,559           &                               & NO1-NO2          & 36,399          \\
Lignite             & 140,826           &                               & NO1-NO3          & 56,874          \\
CCGT                & 70,413            &                               & NO1-NO5          & 52,324          \\
Other gas           & 38,407            &                               & NO2-NO5          & 53,461          \\
Solar               & 66,224            &                               & NO5-NO3          & 79,623          \\
Wind                & 78,735            &                               & NO3-NO4          & 130,809         \\
                    &                   &                               & NO2-DE           & 70,864          \\ \bottomrule
\end{tabular}
\end{table}

Initial generation and transmission capacities are based on data from the EMPIRE model in \cite{backe2022empire}, the ENTSO-E Transparicy platform
\cite{EntsoE-Capacities} and from privately communicated data from THEMA Consulting Group. In line with policy goals of phasing out or reducing production from nuclear plants, some nations have their nuclear plant capacities reduced or completely removed. Germany and Belgium have their nuclear capcities completely removed, while France have a reduced capacity in order to mimic normal operating conditions as well as future capacity goals. Using plant efficiencies, together with privately communicated fuel- and $\mathrm{CO_2}$-price time series from THEMA Consulting Group, we have approximated seasonal marginal operational costs for the thermal power plants. These costs are given in Table \ref{tab:MarginalCostGen}.

\begin{table}[htbp] 
\centering
\caption{Marginal costs [\SI{}{\text{\EUR{}}\per\mega\watt}] for generating technologies.}
\label{tab:MarginalCostGen}
\begin{tabular}{@{}lrrrr@{}}
\toprule
 \multicolumn{1}{c}{\textbf{}}& \multicolumn{1}{c}{\textbf{Winter}}&\multicolumn{1}{c}{\textbf{Spring}}&\multicolumn{1}{c}{\textbf{Summer}}&\multicolumn{1}{c}{\textbf{Autumn}}
  \\ \midrule
Coal    & 35,2 & 29,8 & 30,1 & 35,4 \\
Lignite & 38,9 & 32,9 & 33,3 & 39,4 \\
CCGT    & 41   & 33,8 & 35,2 & 39,3 \\
Gas     & 63,1 & 52,1 & 54,2 & 60,4 \\
Nuclear & 15   & 15   & 15   & 15   \\
Hydro   & 0    & 0    & 0    & 0   \\ \bottomrule
\end{tabular}
\end{table}

Scenario-specific data are sampled from time series from NordPool \cite{Nordpool2022Historical} and Open Power Systems Data \cite{openpowersystemsdata2022timeseries}. The linear demand curves are constructed from demand- and price data. Assuming that inverse demand is written as $\pi = ad + b$, the parameters are calculated as
\begin{align*}
    a = \frac{1}{\varepsilon} \frac{|P|}{D}, \qquad b = \left(1 - \frac{1}{\varepsilon}\right) |P|,
\end{align*}
with $P$ and $D$ being the historical price and demand for a particular hour, respectively, and a price elasticity of demand of $\varepsilon = -0.05$, which is in line with estimates in, e.g., \cite{matar2018households}. We use the absolute value $|P|$ to account for historical hours with negative prices. This method ensures that demand is downward sloping with an elasticity close to \SI{-0.05}{} in cases when model outcomes are close to historical outcomes. Production profiles for solar and wind plants scaled as a factor between 0 and 1 are found from \cite{RenewablesNinja}.

The problem instance consists of a total of 30 scenarios. Each scenario is sampled randomly with historical data from the years 2013-2017. We sample hourly data from consecutive one-week periods, one for each of the four seasons. In this way, each scenario gets four 168-hour seasons resulting in a total of 672 hours in each scenario.

\end{appendices}

\bibliography{references}

\end{document}